\documentclass[leqno,fleqn,11pt]{article}%
\usepackage{MathSpad,makeidx,ifthen,latexsym}
 \usepackage{a4wide,amssymb,amsmath,euler,beton,graphicx}

\newcommand{\flagTikZ}{0}
\newcommand{\omittik}[1]{%
  \ifthenelse{\flagTikZ=1}{#1}{}}
\IfFileExists{tikzflag.tex}{\renewcommand{\flagTikZ}{1}}
 
\omittik{
\usepackage{tikz}
\usetikzlibrary{arrows,automata}
}

\usepackage{hyperref}

\def\mathit#1{#1}

\DeclareSymbolFont{cmsygroup}{OMS}{cmsy}{m}{n}
\DeclareSymbolFont{cmrgroup}{OT1}{cmr}{m}{n}
\DeclareMathSymbol{\sim}{\mathrel}{cmsygroup}{"18}
\DeclareMathSymbol{=}{\mathrel}{cmrgroup}{"3D}
\DeclareMathSymbol{=}{\mathrel}{cmrgroup}{"3D}
\DeclareMathSymbol{[}{\mathopen}{cmrgroup}{"5B}
\DeclareMathSymbol{]}{\mathclose}{cmrgroup}{"5D}
\DeclareMathSymbol{(}{\mathopen}{cmrgroup}{"28}
\DeclareMathSymbol{)}{\mathclose}{cmrgroup}{"29}

\begin{document}
\title{The Thins Ordering on Relations}
\author{Ed Voermans\footnote{Independent researcher, The Netherlands}, 
Jules Desharnais\footnote{Universit\a'{e} Laval, Qu\a'{e}bec, Canada}  and 
Roland Backhouse\footnote{University of Nottingham, UK;
corresponding author}}
\date{\today}
\maketitle
\begin{abstract}Earlier papers \cite{VB2022,VB2023a,VB2023b}  introduced the notions of a core and an index of a
relation (an index being a special case of a core).  A limited form of the axiom of choice was postulated
---specifically that all partial equivalence relations (pers) have an index--- and  the consequences of
adding the axiom to  axiom systems for point-free reasoning were explored.  In this paper, we
define a partial ordering on relations,  which we call the \textsf{thins} ordering.  We show that our axiom of
choice is equivalent to the property that core relations are the minimal elements of the \textsf{thins} ordering. 
We also characterise the relations that are maximal with respect to the \textsf{thins} ordering. Apart from our
axiom of choice, the axiom system we employ is paired to a bare minimum  and  admits many models other than concrete
relations  --- we do not assume, for example, the existence of complements;   in the case of  concrete relations, the theorem is that  the maximal elements of the 
  \textsf{thins}  ordering are the empty relation and the equivalence relations.  
This and other properties of  \textsf{thins}   provide further evidence that our  axiom of choice is
a desirable  means of strengthening  point-free reasoning on relations.\end{abstract}

\section{Introduction}\label{Itt:Introduction}

Earlier papers \cite{VB2022,VB2023a,VB2023b} introduced the notions of a core and an index of a relation (an index
being a special case of a core).    In \cite{VB2023a} the focus was on strengthening standard axiom systems
for point-free reasoning.   A limited form of the axiom of choice was postulated ---specifically that all
partial equivalence relations (pers) have an index--- and  the consequences of adding the axiom were explored. 
The  working document  \cite{VB2022} extends this work to  practical applications of the notions, an extract
of which being \cite{VB2023b} on diagonals and  block-ordered relations.

In this paper, we  define a partial ordering on relations,  which we call the \textsf{thins} ordering.  We begin by
defining \textsf{thins} on partial equivalence relations (pers),  and then extend the ordering to all relations.  We
show that our axiom of choice is equivalent to the property that the minimal elements of the \textsf{thins}
relation on pers are precisely the indexes of pers.  (See theorem  \ref{choice.defs.minimal}  for a precise
statement.)  We then extend the \textsf{thins} ordering to all relations and we show that, assuming our axiom
of choice, the minimal elements of the ordering are precisely the core relations.  (See theorem 
\ref{minimal.is.core}.)  We also show that, when the \textsf{thins} relation  is  restricted to pers, equivalence relations 
are  maximal.   Our calculations make use of a limited subset of the axioms of point-free relation algebra
---for example, we do not assume the existence of complements--- thus testifying to the power of our
axiom of choice.

Because this paper is an extension of  \cite{VB2023a} we have omitted all  introductory material.    For ease
of reference, we do repeat some key topics  from \cite{VB2023a}. In such cases, we omit proofs of lemmas
and theorems.  Hints in our calculations often refer to properties proved in earlier publications; in such
cases we state the properties within square brackets.  (See for example the proof of lemma 
\ref{itt.ABA} where the hint in the first step is \begin{displaymath}\left[\ms{3}\setms{0.15em}P{\MPperdomain}\ms{1}{\MPcomp}\ms{1}P\ms{2}{=}\ms{2}P\ms{2}{=}\ms{2}P\ms{1}{\MPcomp}\ms{1}P{\MPperdomain}\ms{3}\right]~~.\end{displaymath}The square brackets should be read as ``everywhere''.  So the stated property is true for all instances of
the dummy $P$, which ranges in this case over pers.)  Nevertheless, 
  \cite{VB2023a}  is recommended reading   before embarking on the current paper.

Section \ref{Typed Point-Free Relation Algebras} gives a brief summary of our axiom system.   The novel
contributions of the paper begin  in section \ref{Itt:Basic Definitions} with the definition of the \textsf{thins} relation. 
(At this stage, we don't call it an ``ordering'' because that property  has yet to be established.)  We also
reproduce the definition of an index and the axiom of choice from \cite{VB2023a}.  

Section \ref{Itt:Basic Properties} formulates a number of properties of \textsf{thins}.  An important property
(specifically, theorem \ref{corefl.and.thins}) is that the indexes of a per $Q$ are the pers $P$ that thin $Q$ and are
coreflexive.  This section also includes the proof that the \textsf{thins} relation is an ordering relation on pers.

Section \ref{Itt:Minimal and Maximal Pers} is   about pers that are minimal with respect to the
\textsf{thins} ordering.  We show that coreflexive relations  are minimal.  The converse property ---minimal
implies coreflexive--- is then shown to be equivalent to our axiom of choice.  The conclusion of the
section, theorem \ref{choice.defs.minimal}, is that the axiom of choice is equivalent  to the conjunction of two
properties:  firstly, the minimal elements of the \textsf{thins} ordering  on pers are precisely the coreflexive relations and,
secondly, every per thins to a minimal element.    


Section \ref{Itt:Maximality} is about maximality.  A brief, informal summary of the main theorem (theorem 
\ref{maximal.char})  is  that the equivalence relations are maximal with respect to the \textsf{thins} ordering on
non-empty pers.  The theorem we formulate is, in fact, more general than this since it applies to models
of point-free relation algebra quite different from the standard set-theoretic binary relations.   The
statement of the theorem introduces a new idiom to point-free relation algebra that avoids a case
analysis on whether or not a relation is empty.

Section \ref{General.thins.relation} is where we extend the \textsf{thins} ordering to arbitrary relations (and not just
pers).  We prove that, assuming our axiom of choice, a relation $S$ is minimal with respect to the \textsf{thins} ordering 
on arbitrary  relations iff $S$ is a core relation.  



\section{Point-Free Relation Algebras}\label{Typed Point-Free Relation Algebras}

In this section, we define a \emph{point-free relation algebra}.     Such an
algebra has three  components with interfaces between them:  a (typed) monoid structure,
a  lattice structure,  and a converse structure.  

Underpinning any relation algebra is a very simple type structure.   We assume the existence of a 
non-empty set $\mathcal{T}$   of so-called \emph{basic types}.   A \emph{relation type} is an ordered  pair of basic types.  We write
$A\ms{1}{\sim}\ms{1}B$  for the ordered pair of basic types $A$ and $B$.  We often omit ``relation'' and refer to $A\ms{1}{\sim}\ms{1}B$ as a ``type''.
 The carrier set of a point-free relation  algebra is typed in the sense that each element $X$ of the carrier
set has a type $A\ms{1}{\sim}\ms{1}B$ for some basic types $A$ and $B$.  A relation of type $A\ms{1}{\sim}\ms{1}A$, for some $A$, is said to be
\emph{homogeneous}.   If $\mathcal{T}$ has exactly one element we say that the algebra is   \emph{untyped}.   

The monoid structure is defined as follows.  For each triple of basic types $A$, $B$ and $C$, and each element  $X$ of
type  $A\ms{1}{\sim}\ms{1}B$ and each element  $Y$ of type $B\ms{1}{\sim}\ms{1}C$, there is an element  $X{\MPcomp}Y$ of type $A\ms{1}{\sim}\ms{1}C$.  Also for each basic type $A$
there is an element  $\mathbb{I}$ of type $A\ms{1}{\sim}\ms{1}A$.  The element $X{\MPcomp}Y$ is called the \emph{composition} of $X$ and $Y$, and  $\mathbb{I}$ is called the
\emph{identity} of $A$.    Composition is required to be associative, and identities are required to be the units of
composition.  The composition $X{\MPcomp}Y$ is only defined when $X$ and $Y$ have appropriate types.  (Such a typed
monoid structure is commonly called a ``category''.)

In principle,  the type of the identities should be made explicit in the notation we use: for example, by
writing $\mathbb{I}_{A}$ for the identity of type $A\ms{1}{\sim}\ms{1}A$.  It is convenient for us not to do so, leaving the type information
to be deduced from the context.  This is also the case for other operators and constants that we introduce
below.

For each type $A\ms{1}{\sim}\ms{1}B$,  we assume the existence of  a (finitely) distributive lattice  (partially)  ordered by ${\subseteq}$ . 
The binary supremum and infimum operators of the lattice are denoted by ${\cup}$  and ${\cap}$, respectively.    The
least and greatest elements of the lattice are denoted by  ${\MPplatbottom}$ and ${\MPplattop}$ , respectively.  

The interface between the monoid structure and the lattice structure is the existence of the two factor 
operators defined by, for all $X$, $Y$ and $Z$ of  appropriate types,\begin{displaymath}(Y\ms{2}{\subseteq}\ms{2}X{\setminus}Z)\ms{4}{=}\ms{4}(X{\MPcomp}Y\ms{2}{\subseteq}\ms{2}Z)\ms{4}{=}\ms{4}(X\ms{2}{\subseteq}\ms{2}Z{/}Y)~~.\end{displaymath}As a consequence,  composition distributes over supremum.  
That is,  for all relations $R$, $S$, $T$  and $U$ of appropriate  type\begin{displaymath}R{\MPcomp}(S{\cup}T)\ms{4}{=}\ms{4}R{\MPcomp}S\ms{1}{\cup}\ms{1}R{\MPcomp}T~~,\end{displaymath} \begin{displaymath}(S{\cup}T){\MPcomp}U\ms{4}{=}\ms{4}S{\MPcomp}U\ms{1}{\cup}\ms{1}T{\MPcomp}U~~,\end{displaymath}\begin{displaymath}R{\MPcomp}{\MPplatbottom}\ms{3}{=}\ms{3}{\MPplatbottom}\mbox{~~, and}\end{displaymath}\begin{displaymath}{\MPplatbottom}{\MPcomp}R\ms{3}{=}\ms{3}{\MPplatbottom}~~.\end{displaymath}(The symbol ${\MPplatbottom}$ is overloaded in the final two equations:  each occurrence may have a different type.)

The converse structure is very simple: for each element  $X$  of type $A\ms{1}{\sim}\ms{1}B$ there is an element
$X^{\MPrev}$ of type $B\ms{1}{\sim}\ms{1}A$.  

The interface between the lattice structure and the converse structure is the Galois connection: for all $X$
and $Y$ of appropriate types,\begin{displaymath}X^{\MPrev}\ms{2}{\subseteq}\ms{2}Y\ms{6}{\equiv}\ms{6}X\ms{2}{\subseteq}\ms{2}Y^{\MPrev}~~.\end{displaymath}The interface between the monoid structure and the converse structure is:  for each identity $\mathbb{I}$, \begin{displaymath}\mathbb{I}^{\MPrev}\ms{2}{=}\ms{2}\mathbb{I}\end{displaymath}and, for all $X$ and $Y$ of appropriate types,  \begin{displaymath}(X{\MPcomp}Y)^{\MPrev}\ms{4}{=}\ms{4}Y^{\MPrev}\ms{1}{\MPcomp}\ms{1}X^{\MPrev}~~.\end{displaymath}Finally, the \emph{modularity law} acts as an interface between all three components: for all $X$, $Y$ and $Z$ of
appropriate types,\begin{displaymath}X{\MPcomp}Y\ms{2}{\cap}\ms{2}Z\ms{3}{\subseteq}\ms{3}X\ms{1}{\MPcomp}\ms{1}(Y\ms{3}{\cap}\ms{3}X^{\MPrev}\ms{1}{\MPcomp}\ms{1}Z)~~.\end{displaymath}We do not use the existence of the factor operators or the 
modularity law anywhere explicitly in this paper.  We do, however, make
extensive use of the properties of coreflexive relations and 
the (coreflexive-)domain operators first mentioned in section \ref{Itt:Basic Definitions}, as well as the
per-domain operators in section \ref{General.thins.relation}.  The  properties  of coreflexive domains 
rely heavily on the  modularity law, and per domains are defined in terms of the factor operators.    

A consequence of the axioms is that all the operators of the  algebra ---composition,
converse, supremum,  infimum and the domain operators introduced later--- are monotonic with
respect to the ${\subseteq}$ ordering.  We exploit monotonicity frequently in our calculations, sometimes without
explicit mention.

The axioms of point-free relation algebra do not completely characterise all the properties of
binary relations and, therefore, admit other models (for example geometric models: see \cite[2.158]{FRSC90} and 
\cite[section 3.5]{Vo99}).  We use the term \emph{concrete relation} below to refer to  binary relations as they are
normally understood.  That is, a \emph{concrete relation of type}  $A\ms{1}{\sim}\ms{1}B$ is an element of the powerset $2^{A{\MPtimes}B}$.  
(The types $A$ and $B$ do not need to be finite.)

A point-free relation  algebra is said to be  \emph{unary}  if it satisfies the \emph{cone rule}: for all $X$, \begin{displaymath}{\MPplattop}{\MPcomp}X{\MPcomp}{\MPplattop}\ms{1}{=}\ms{1}{\MPplattop}\ms{6}{\equiv}\ms{6}X\ms{1}{\neq}\ms{1}{\MPplatbottom}~~.\end{displaymath}(The three occurrences of ${\MPplattop}$ may have different types.  The rightmost occurrence is assumed to have the
same type as $X$; the other two are assumed to be homogeneous relations of the appropriate types.
The terminology reflects the fact that the cartesian product of two relation algebras  is non-unary.  
See \cite[section 3.4.3]{Vo99}.)

In  \cite{VB2022,VB2023a}, much of the focus was on introducing axioms that facilitate pointwise reasoning.
To this end,  the cone rule was used extensively.  In contrast, in this paper the goal is not to facilitate
pointwise reasoning but, instead, to strengthen point-free reasoning.  So here the cone rule is deemed 
to be invalid.  See  the introductory remarks  in section \ref{Itt:Maximality}.  
\begin{Example}\label{simplest.point-free.algebras}{\rm \ \ \ The simplest examples of point-free relation algebras are all 
untyped.  The simplest of all has just one element:  all of the constants ${\MPplatbottom}$,   $\mathbb{I}$ and ${\MPplattop}$ are defined to be equal. 
The second simplest has two elements ${\MPplatbottom}$ and ${\MPplattop}$;  $\mathbb{I}$ is defined to be equal to ${\MPplattop}$ (and different from ${\MPplatbottom}$).  The
third simplest  has three elements: the constants ${\MPplatbottom}$, $\mathbb{I}$ and ${\MPplattop}$, which are defined to be distinct.   (In all three
cases,  the definitions of the ordering relation, composition and converse can be deduced from the 
axioms.)  

A  four-element algebra  is obtained by adding a new  element ${\neg}\mathbb{I}$ to the three-element algebra 
and defining the
composition ${\neg}\mathbb{I}\ms{1}{\MPcomp}\ms{1}{\neg}\mathbb{I}$ to be $\mathbb{I}$ and the  converse $({\neg}\mathbb{I})^{\MPrev}$ to be ${\neg}\mathbb{I}$.  As suggested by the notation,  ${\neg}\mathbb{I}$ is 
the complement of $\mathbb{I}$.  That is,  the lattice structure is as shown in the diagram below.

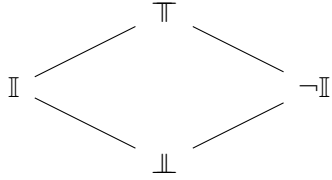
\omittik{

\begin{center}
\begin{tikzpicture}[>=stealth,initial text=]
\node (1) at (0,2) {$\MPplattop$};
\node (2) at (-2,1) {$\mathbb{I}$};
\node (3) at (2,1) {$\neg\mathbb{I}$};
\node (4) at (0,0) {$\MPplatbottom$};
\path[-]
(1) edge node {} (2)
(1) edge node {} (3)
(2) edge node {} (4)
(3) edge node {} (4)
;
\end{tikzpicture}
\end{center}

}

The simplest example is not unary, the other  examples are unary.    A model of the two-element algebra
is formed by the (homogeneous) concrete  relations on a set with exactly one element.  The other 
 examples do not have such a model since the concrete  relations on a set
of size $n$  form a power set of size $2^{n{\times}n}$.
}%
\MPendBox\end{Example}

\section{Basic Definitions}\label{Itt:Basic Definitions}

We begin by restricting our study to partial equivalence relations (pers\footnote{Relation  $P$  is a per iff it
is  symmetric  (i.e.\ $P\ms{1}{=}\ms{1}P^{\MPrev}$) and transitive (i.e.\ $P{\MPcomp}P\ms{1}{\subseteq}\ms{1}P$). Equivalently,  $P$ is a per iff $P\ms{2}{=}\ms{2}P\ms{1}{\MPcomp}\ms{1}P^{\MPrev}$.}). 
In this section we recall the definition of an
index of a per and our axiom of choice.  New is definition \ref{itt.def}.

Throughout the paper,  $P$ and $Q$ denote pers.  For pers,  the left and right domains coincide.  
(I.e. for all pers $P$,  $P{\MPldom{}}\ms{1}{=}\ms{1}P{\MPrdom{}}$.)  For this reason, $P{\MPperdomain}$ is used to denote the left/right domain of $P$.  
That is,  $P{\MPldom{}}\ms{1}{=}\ms{1}P{\MPperdomain}\ms{1}{=}\ms{1}P{\MPrdom{}}$.  (We assume familiarity with the properties of coreflexive\footnote{Relation $p$ is
coreflexive iff  $p\ms{1}{\subseteq}\ms{1}\mathbb{I}$.} relations and the domain
operators.  So, rather than include an extensive list of their properties,  we state the properties being used
between square ``everywhere'' brackets, as explained earlier.)
\begin{Definition}[Index of a Per]\label{per.index}{\rm \ \ \ Suppose $P$ is a per.  Then an 
 \emph{index} of $P$ is a relation $J$ such  that 
\begin{description}
\item[(a)]$J\ms{1}{\subseteq}\ms{1}P{\MPperdomain}~~,$ 
\item[(b)]$J{\MPcomp}P{\MPcomp}J\ms{2}{=}\ms{2}J~~,$ 
\item[(c)]$P{\MPcomp}J{\MPcomp}P\ms{2}{=}\ms{2}P~~.$ 
 
\end{description}
\vspace{-7mm}
}
\MPendBox\end{Definition}
\begin{Axiom}[Axiom of Choice]\label{Axiom of Choice}{\rm \ \ \ Every per has an index.
}
\MPendBox\end{Axiom}

\begin{Example}\label{counterexample.axiom.of.choice}{\rm \ \ \ The three- and four-element algebras detailed in example 
\ref{simplest.point-free.algebras} do not  satisfy the axiom of choice since, 
in both  cases,  ${\MPplattop}$ does not have an  index.  The two simplest examples do satisfy the axiom of choice because
each element is an index of itself.
}%
\MPendBox\end{Example}
\begin{Definition}[Thins]\label{itt.def}{\rm \ \ \ The \textsf{thins} relation on pers is defined by, for all pers $P$ and $Q$, \begin{displaymath}P\ms{1}{\preceq}\ms{1}Q\ms{8}{\equiv}\ms{8}P\ms{2}{=}\ms{2}P{\MPperdomain}\ms{1}{\MPcomp}\ms{1}Q\ms{1}{\MPcomp}\ms{1}P{\MPperdomain}\ms{5}{\wedge}\ms{5}Q\ms{2}{=}\ms{2}Q\ms{1}{\MPcomp}\ms{1}P{\MPperdomain}\ms{1}{\MPcomp}\ms{1}Q~~.\end{displaymath}\vspace{-9mm}
}
\MPendBox\end{Definition}

This paper is about the properties of the ${\preceq}$ relation.  We call it the \textsf{thins} relation.  (So $P\ms{1}{\preceq}\ms{1}Q$ is
pronounced $P$ \textsf{thins} $Q$.) Much of the paper is about the \textsf{thins} relation on pers but we extend it to all
relations in section \ref{General.thins.relation}. 

Informally, the first conjunct in the definition of $P\ms{1}{\preceq}\ms{1}Q$ states that the equivalence classes of $P$ are subsets
of the equivalence classes of $Q$, and the second conjunct states that,  for each equivalence class of $P$, there is 
a corresponding  equivalence class  of  $Q$.  Some  of the properties stated below are intended to confirm
this informal  interpretation of the definition.   
\begin{Example}\label{example.thins}{\rm \ \ \ In all four example algebras detailed in example 
\ref{simplest.point-free.algebras}, the pers  are ${\MPplatbottom}$,  $\mathbb{I}$ and ${\MPplattop}$  and the  \textsf{thins} relation is discrete.  
(That is, the \textsf{thins} relation is the equality relation on the pers.)
}%
\MPendBox\end{Example}

\section{Basic Properties}\label{Itt:Basic Properties}
 
As the title suggests, this section is about basic properties of the \textsf{thins} relation.   Theorem \ref{itt.order}
establishes that it is a partial ordering on pers.  Theorem \ref{corefl.and.thins} formulates an alternative
definition of an index of a per in terms of \textsf{thins}.  Subsequent lemmas anticipate properties needed in
later sections.

Obvious from the property that $P{\MPperdomain}\ms{1}{\subseteq}\ms{1}\mathbb{I}$  (for arbitrary $P$) and monotonicity of composition, applied to 
the property  $P\ms{2}{=}\ms{2}P{\MPperdomain}\ms{1}{\MPcomp}\ms{1}Q\ms{1}{\MPcomp}\ms{1}P{\MPperdomain}$  is that, for all pers $P$ and $Q$,\begin{equation}\label{itt.imp.atmost}
P\ms{1}{\preceq}\ms{1}Q\ms{4}{\Rightarrow}\ms{4}P\ms{1}{\subseteq}\ms{1}Q~~.
\end{equation}We use this frequently below.  

The following lemma is also used on several occasions.   Compare the lemma with properties 
\ref{per.index}(b) and \ref{per.index}(c) of an index.
\begin{Lemma}\label{itt.ABA}{\rm \ \ \ For all pers $P$ and  $Q$,\begin{displaymath}P\ms{1}{\preceq}\ms{1}Q\ms{8}{\Rightarrow}\ms{8}P\ms{2}{=}\ms{2}P{\MPcomp}Q{\MPcomp}P\ms{4}{\wedge}\ms{4}Q\ms{2}{=}\ms{2}Q{\MPcomp}P{\MPcomp}Q~~.\end{displaymath}
}%
\end{Lemma}%
{\bf Proof}~~~Assume $P\ms{2}{=}\ms{2}P{\MPperdomain}\ms{1}{\MPcomp}\ms{1}Q\ms{1}{\MPcomp}\ms{1}P{\MPperdomain}$.  Then
\begin{mpdisplay}{0.15em}{6.5mm}{0mm}{2}
	$P{\MPcomp}Q{\MPcomp}P$\push\-\\
	$=$	\>	\>$\{$	\>\+\+\+domains (specifically     $\left[\ms{3}P{\MPperdomain}\ms{1}{\MPcomp}\ms{1}P\ms{2}{=}\ms{2}P\ms{2}{=}\ms{2}P\ms{1}{\MPcomp}\ms{1}P{\MPperdomain}\ms{3}\right]$)\-\-$~~~ \}$\pop\\
	$P\ms{1}{\MPcomp}\ms{1}P{\MPperdomain}\ms{1}{\MPcomp}\ms{1}Q\ms{1}{\MPcomp}\ms{1}P{\MPperdomain}\ms{1}{\MPcomp}\ms{1}P$\push\-\\
	$=$	\>	\>$\{$	\>\+\+\+assumption:  $P\ms{2}{=}\ms{2}P{\MPperdomain}\ms{1}{\MPcomp}\ms{1}Q\ms{1}{\MPcomp}\ms{1}P{\MPperdomain}$\-\-$~~~ \}$\pop\\
	$P{\MPcomp}P{\MPcomp}P$\push\-\\
	$=$	\>	\>$\{$	\>\+\+\+$P$ is a per,   $\left[\ms{2}P\ms{1}{=}\ms{1}P{\MPcomp}P\ms{2}\right]$\-\-$~~~ \}$\pop\\
	$P~~.$
\end{mpdisplay}
Since $P\ms{1}{\preceq}\ms{1}Q\ms{6}{\Rightarrow}\ms{6}P\ms{2}{=}\ms{2}P{\MPperdomain}\ms{1}{\MPcomp}\ms{1}Q\ms{1}{\MPcomp}\ms{1}P{\MPperdomain}$,  it follows that $P\ms{1}{\preceq}\ms{1}Q\ms{5}{\Rightarrow}\ms{5}P\ms{2}{=}\ms{2}P{\MPcomp}Q{\MPcomp}P$.  Also,
\begin{mpdisplay}{0.15em}{6.5mm}{0mm}{2}
	$Q{\MPcomp}P{\MPcomp}Q$\push\-\\
	$\supseteq$	\>	\>$\{$	\>\+\+\+$\left[\ms{2}P\ms{1}{\supseteq}\ms{1}P{\MPperdomain}\ms{2}\right]$\-\-$~~~ \}$\pop\\
	$Q\ms{1}{\MPcomp}\ms{1}P{\MPperdomain}\ms{1}{\MPcomp}\ms{1}Q$\push\-\\
	$=$	\>	\>$\{$	\>\+\+\+assumption: $P\ms{1}{\preceq}\ms{1}Q$,  so  $Q\ms{2}{=}\ms{2}Q\ms{1}{\MPcomp}\ms{1}P{\MPperdomain}\ms{1}{\MPcomp}\ms{1}Q$\-\-$~~~ \}$\pop\\
	$Q$\push\-\\
	$=$	\>	\>$\{$	\>\+\+\+ $Q$ is a per,    $\left[\ms{2}P\ms{1}{=}\ms{1}P{\MPcomp}P\ms{2}\right]$ with $P\ms{1}{:=}\ms{1}Q$\-\-$~~~ \}$\pop\\
	$Q{\MPcomp}Q{\MPcomp}Q$\push\-\\
	$\supseteq$	\>	\>$\{$	\>\+\+\+(\ref{itt.imp.atmost}) and monotonicity\-\-$~~~ \}$\pop\\
	$Q{\MPcomp}P{\MPcomp}Q~~.$
\end{mpdisplay}
That is,  by anti-symmetry,   $P\ms{1}{\preceq}\ms{1}Q\ms{5}{\Rightarrow}\ms{5}Q\ms{2}{=}\ms{2}Q{\MPcomp}P{\MPcomp}Q~~.$
\MPendBox

The notation we have chosen suggests that \textsf{thins} is a partial ordering.  This is indeed the case:
\begin{Theorem}\label{itt.order}{\rm \ \ \ The \textsf{thins} relation is a partial ordering on pers.
}
\end{Theorem}
{\bf Proof}~~~We must prove that the \textsf{thins} relation  is reflexive, transitive and anti-symmetric.   
Reflexivity is straightforward:
\begin{mpdisplay}{0.15em}{6.5mm}{0mm}{2}
	$P\ms{1}{\preceq}\ms{1}P$\push\-\\
	$=$	\>	\>$\{$	\>\+\+\+definition \ref{itt.def}\-\-$~~~ \}$\pop\\
	$P\ms{2}{=}\ms{2}P{\MPperdomain}\ms{1}{\MPcomp}\ms{1}P\ms{1}{\MPcomp}\ms{1}P{\MPperdomain}\ms{5}{\wedge}\ms{5}P\ms{2}{=}\ms{2}P\ms{1}{\MPcomp}\ms{1}P{\MPperdomain}\ms{1}{\MPcomp}\ms{1}P$\push\-\\
	$=$	\>	\>$\{$	\>\+\+\+domains (specifically,  $\left[\ms{3}P{\MPperdomain}\ms{1}{\MPcomp}\ms{1}P\ms{2}{=}\ms{2}P\ms{2}{=}\ms{2}P\ms{1}{\MPcomp}\ms{1}P{\MPperdomain}\ms{3}\right]$)\-\-$~~~ \}$\pop\\
	$P\ms{1}{=}\ms{1}P\ms{4}{\wedge}\ms{4}P\ms{1}{=}\ms{1}P{\MPcomp}P$\push\-\\
	$=$	\>	\>$\{$	\>\+\+\+reflexivitiy of equality;  $P$ is a per and  $\left[\ms{2}P\ms{1}{=}\ms{1}P{\MPcomp}P\ms{2}\right]$\-\-$~~~ \}$\pop\\
	$\mathsf{true}~~.$
\end{mpdisplay}
Now, suppose $P$, $Q$   and $R$ are pers, and  $P\ms{1}{\preceq}\ms{1}Q$ and $Q\ms{1}{\preceq}\ms{1}R$.  Applying (\ref{itt.imp.atmost}), we have:\begin{equation}\label{ABC}
P\ms{1}{\subseteq}\ms{1}Q\ms{1}{\subseteq}\ms{1}R~~.
\end{equation}To prove transitivity, we must prove that $P\ms{1}{\preceq}\ms{1}R$.  Applying definition \ref{itt.def}, we must prove that \begin{equation}\label{ACA}
P\ms{3}{=}\ms{3}P{\MPperdomain}\ms{1}{\MPcomp}\ms{1}R\ms{1}{\MPcomp}\ms{1}P{\MPperdomain}
\end{equation}and \begin{equation}\label{CAC}
R\ms{3}{=}\ms{3}R\ms{1}{\MPcomp}\ms{1}P{\MPperdomain}\ms{1}{\MPcomp}\ms{1}R~~.
\end{equation}We prove (\ref{ACA})  by mutual inclusion: 
\begin{mpdisplay}{0.15em}{6.5mm}{0mm}{2}
	$P$\push\-\\
	$=$	\>	\>$\{$	\>\+\+\+domains (specifically,  $\left[\ms{3}P{\MPperdomain}\ms{1}{\MPcomp}\ms{1}P\ms{2}{=}\ms{2}P\ms{2}{=}\ms{2}P\ms{1}{\MPcomp}\ms{1}P{\MPperdomain}\ms{3}\right]$)\-\-$~~~ \}$\pop\\
	$P{\MPperdomain}\ms{1}{\MPcomp}\ms{1}P\ms{1}{\MPcomp}\ms{1}P{\MPperdomain}$\push\-\\
	$\subseteq$	\>	\>$\{$	\>\+\+\+by (\ref{ABC}),  $P\ms{1}{\subseteq}\ms{1}R$~; monotonicity\-\-$~~~ \}$\pop\\
	$P{\MPperdomain}\ms{1}{\MPcomp}\ms{1}R\ms{1}{\MPcomp}\ms{1}P{\MPperdomain}$\push\-\\
	$\subseteq$	\>	\>$\{$	\>\+\+\+$P$ is a per and   $\left[\ms{2}P{\MPperdomain}\ms{1}{\subseteq}\ms{1}P\ms{2}\right]$~; monotonicity\-\-$~~~ \}$\pop\\
	$P\ms{1}{\MPcomp}\ms{1}R\ms{1}{\MPcomp}\ms{1}P$\push\-\\
	$=$	\>	\>$\{$	\>\+\+\+assumption:  $P\ms{2}{=}\ms{2}P{\MPperdomain}\ms{1}{\MPcomp}\ms{1}Q\ms{1}{\MPcomp}\ms{1}P{\MPperdomain}$\-\-$~~~ \}$\pop\\
	$P{\MPperdomain}\ms{2}{\MPcomp}\ms{2}Q\ms{2}{\MPcomp}\ms{2}P{\MPperdomain}\ms{2}{\MPcomp}\ms{2}R\ms{2}{\MPcomp}\ms{2}P{\MPperdomain}\ms{2}{\MPcomp}\ms{2}Q\ms{2}{\MPcomp}\ms{2}P{\MPperdomain}$\push\-\\
	$\subseteq$	\>	\>$\{$	\>\+\+\+$\left[\ms{2}P{\MPperdomain}\ms{1}{\subseteq}\ms{1}\mathbb{I}\ms{2}\right]$; monotonicity\-\-$~~~ \}$\pop\\
	$P{\MPperdomain}\ms{2}{\MPcomp}\ms{2}Q\ms{2}{\MPcomp}\ms{2}R\ms{2}{\MPcomp}\ms{2}Q\ms{2}{\MPcomp}\ms{2}P{\MPperdomain}$\push\-\\
	$=$	\>	\>$\{$	\>\+\+\+assumption: $Q\ms{1}{\preceq}\ms{1}R$,   lemma \ref{itt.ABA} with $P{,}Q\ms{1}{:=}\ms{1}Q{,}R$\-\-$~~~ \}$\pop\\
	$P{\MPperdomain}\ms{2}{\MPcomp}\ms{2}Q\ms{2}{\MPcomp}\ms{2}P{\MPperdomain}$\push\-\\
	$=$	\>	\>$\{$	\>\+\+\+assumption:  $P\ms{2}{=}\ms{2}P{\MPperdomain}\ms{1}{\MPcomp}\ms{1}Q\ms{1}{\MPcomp}\ms{1}P{\MPperdomain}$\-\-$~~~ \}$\pop\\
	$P~~.$
\end{mpdisplay}
Now we prove (\ref{CAC}).  Again, the proof is by mutual inclusion:
\begin{mpdisplay}{0.15em}{6.5mm}{0mm}{2}
	$R\ms{1}{\MPcomp}\ms{1}P{\MPperdomain}\ms{1}{\MPcomp}\ms{1}R$\push\-\\
	$\subseteq$	\>	\>$\{$	\>\+\+\+$\left[\ms{2}P{\MPperdomain}\ms{1}{\subseteq}\ms{1}\mathbb{I}\ms{2}\right]$ and monotonicity\-\-$~~~ \}$\pop\\
	$R{\MPcomp}R$\push\-\\
	$\subseteq$	\>	\>$\{$	\>\+\+\+$R$ is a per (and hence transitive)\-\-$~~~ \}$\pop\\
	$R$\push\-\\
	$=$	\>	\>$\{$	\>\+\+\+assumption:  $Q\ms{1}{\preceq}\ms{1}R$,  lemma \ref{itt.ABA} with $P{,}Q\ms{1}{:=}\ms{1}Q{,}R$\-\-$~~~ \}$\pop\\
	$R{\MPcomp}Q{\MPcomp}R$\push\-\\
	$=$	\>	\>$\{$	\>\+\+\+assumption:  $P\ms{1}{\preceq}\ms{1}Q$\-\-$~~~ \}$\pop\\
	$R\ms{1}{\MPcomp}\ms{1}Q\ms{1}{\MPcomp}\ms{1}P{\MPperdomain}\ms{1}{\MPcomp}\ms{1}Q\ms{1}{\MPcomp}\ms{1}R$\push\-\\
	$\subseteq$	\>	\>$\{$	\>\+\+\+assumption:  $Q\ms{1}{\preceq}\ms{1}R$,  (\ref{itt.imp.atmost}) and monotonicity\-\-$~~~ \}$\pop\\
	$R\ms{1}{\MPcomp}\ms{1}R\ms{1}{\MPcomp}\ms{1}P{\MPperdomain}\ms{1}{\MPcomp}\ms{1}R\ms{1}{\MPcomp}\ms{1}R$\push\-\\
	$\subseteq$	\>	\>$\{$	\>\+\+\+$R$ is a per (and hence transitive)\-\-$~~~ \}$\pop\\
	$R\ms{1}{\MPcomp}\ms{1}P{\MPperdomain}\ms{1}{\MPcomp}\ms{1}R~~.$
\end{mpdisplay}
Finally, combining (\ref{ACA}) and (\ref{CAC}) and applying definition \ref{itt.def} (with $P{,}Q\ms{1}{:=}\ms{1}P{,}R$) we have shown
that $P\ms{1}{\preceq}\ms{1}R$.  This concludes the proof that the \textsf{thins}  relation is transitive.

Finally, we prove that the \textsf{thins}  relation is  anti-symmetric.  Suppose $P\ms{1}{\preceq}\ms{1}Q\ms{1}{\preceq}\ms{1}P$.    Then, by
 (\ref{itt.imp.atmost}),  $P\ms{1}{\subseteq}\ms{1}Q\ms{1}{\subseteq}\ms{1}P$.  Thus $P\ms{1}{=}\ms{1}Q$ by the anti-symmetry of the ${\subseteq}$ relation.
\MPendBox

We now consider the properties of indexes with respect to the \textsf{thins} ordering.
Our first goal is to show that the definition of an index $J$ of a per $P$ can be split into two
conjuncts, namely $J\ms{1}{\subseteq}\ms{1}\mathbb{I}$ and $J\ms{1}{\preceq}\ms{1}P$.    (See theorem \ref{corefl.and.thins}.)  First, a lemma:

\begin{Lemma}\label{index.itt}{\rm \ \ \ If  $J$ is an index of $P$ then  $J\ms{1}{\preceq}\ms{1}P$.
}%
\end{Lemma}%
{\bf Proof}~~~ 
\begin{mpdisplay}{0.15em}{6.5mm}{0mm}{2}
	$J$  is an index of $P$\push\-\\
	$=$	\>	\>$\{$	\>\+\+\+definition \ref{per.index}\-\-$~~~ \}$\pop\\
	$J\ms{1}{\subseteq}\ms{1}P{\MPperdomain}\ms{5}{\wedge}\ms{5}J{\MPcomp}P{\MPcomp}J\ms{2}{=}\ms{2}J\ms{5}{\wedge}\ms{5}P{\MPcomp}J{\MPcomp}P\ms{2}{=}\ms{2}P$\push\-\\
	$\Rightarrow$	\>	\>$\{$	\>\+\+\+domains (specifically,  $\left[\ms{2}P{\MPperdomain}\ms{1}{\subseteq}\ms{1}\mathbb{I}\ms{2}\right]$  and  $\left[\ms{3}Q\ms{1}{\subseteq}\ms{1}\mathbb{I}\ms{2}{\equiv}\ms{2}Q\ms{1}{=}\ms{1}Q{\MPperdomain}\ms{3}\right]$ with $Q\ms{1}{:=}\ms{1}J$)\-\-$~~~ \}$\pop\\
	$J\ms{1}{=}\ms{1}J{\MPperdomain}\ms{5}{\wedge}\ms{5}J{\MPcomp}P{\MPcomp}J\ms{2}{=}\ms{2}J\ms{5}{\wedge}\ms{5}P{\MPcomp}J{\MPcomp}P\ms{2}{=}\ms{2}P$\push\-\\
	$\Rightarrow$	\>	\>$\{$	\>\+\+\+Leibniz\-\-$~~~ \}$\pop\\
	$J{\MPperdomain}\ms{1}{\MPcomp}\ms{1}P\ms{1}{\MPcomp}\ms{1}J{\MPperdomain}\ms{3}{=}\ms{3}J\ms{6}{\wedge}\ms{6}P\ms{1}{\MPcomp}\ms{1}J{\MPperdomain}\ms{1}{\MPcomp}\ms{1}P\ms{3}{=}\ms{3}P$\push\-\\
	$=$	\>	\>$\{$	\>\+\+\+definition \ref{itt.def}\-\-$~~~ \}$\pop\\
	$J\ms{1}{\preceq}\ms{1}P~~.$
\end{mpdisplay}
\vspace{-7mm}
\MPendBox

\begin{Theorem}\label{corefl.and.thins}{\rm \ \ \ For all pers $P$ and  $Q$,\begin{displaymath}P\ms{1}{\subseteq}\ms{1}\mathbb{I}\ms{4}{\wedge}\ms{4}P\ms{1}{\preceq}\ms{1}Q\ms{9}{\equiv}\ms{9}P\mbox{  is an  index of }Q~~.\end{displaymath}
}
\end{Theorem}
{\bf Proof}~~~The proof is by mutual implication.  For ease of reference, we instantiate definition \ref{per.index}
with $J{,}P\ms{1}{:=}\ms{1}P{,}Q$:\begin{equation}\label{corefl.and.thins0}
P\ms{1}{\subseteq}\ms{1}Q{\MPperdomain}\ms{5}{\wedge}\ms{5}P{\MPcomp}Q{\MPcomp}P\ms{2}{=}\ms{2}P\ms{5}{\wedge}\ms{5}Q{\MPcomp}P{\MPcomp}Q\ms{2}{=}\ms{2}Q~~.
\end{equation}Suppose $P\ms{1}{\subseteq}\ms{1}\mathbb{I}\ms{3}{\wedge}\ms{3}P\ms{1}{\preceq}\ms{1}Q$.  We must verify (\ref{corefl.and.thins0}).  The first conjunct is verified as follows:  
\begin{mpdisplay}{0.15em}{6.5mm}{0mm}{2}
	$P\ms{1}{\subseteq}\ms{1}Q{\MPperdomain}$\push\-\\
	$=$	\>	\>$\{$	\>\+\+\+assumption:  $P\ms{1}{\subseteq}\ms{1}\mathbb{I}$~; so $P\ms{1}{=}\ms{1}P{\MPperdomain}$\-\-$~~~ \}$\pop\\
	$P{\MPperdomain}\ms{1}{\subseteq}\ms{1}Q{\MPperdomain}$\push\-\\
	$\Leftarrow$	\>	\>$\{$	\>\+\+\+monotonicity\-\-$~~~ \}$\pop\\
	$P\ms{1}{\subseteq}\ms{1}Q$\push\-\\
	$\Leftarrow$	\>	\>$\{$	\>\+\+\+(\ref{itt.imp.atmost})\-\-$~~~ \}$\pop\\
	$P\ms{1}{\preceq}\ms{1}Q~~.$
\end{mpdisplay}
The second and third conjuncts  both follow directly from the assumption $P\ms{1}{\preceq}\ms{1}Q$ by 
lemma \ref{itt.ABA}.

For the converse implication, we have:
\begin{mpdisplay}{0.15em}{6.5mm}{0mm}{2}
	$P$ is an index of $Q$\push\-\\
	$\Rightarrow$	\>	\>$\{$	\>\+\+\+definition \ref{per.index}(a)  and lemma \ref{index.itt}  (both with $J{,}P\ms{1}{:=}\ms{1}P{,}Q$)\-\-$~~~ \}$\pop\\
	$P\ms{1}{\subseteq}\ms{1}Q{\MPperdomain}\ms{5}{\wedge}\ms{5}P\ms{1}{\preceq}\ms{1}Q$\push\-\\
	$\Rightarrow$	\>	\>$\{$	\>\+\+\+domains (specifically $\left[\ms{2}P{\MPperdomain}\ms{1}{\subseteq}\ms{1}\mathbb{I}\ms{2}\right]$ with $P\ms{1}{:=}\ms{1}Q$) and transitivity of ${\subseteq}$\-\-$~~~ \}$\pop\\
	$P\ms{1}{\subseteq}\ms{1}\mathbb{I}\ms{4}{\wedge}\ms{4}P\ms{1}{\preceq}\ms{1}Q~~.$
\end{mpdisplay}
\vspace{-7mm}
\MPendBox

Case analysis is a commonly used reasoning strategy but it is something we want to avoid whenever
possible.  Case analysis occurs when complements are used:  for example, 
when using the law of the excluded middle.  Using the cone rule also leads to case analysis: on when a
relation is empty or non-empty.  In section \ref{Itt:Maximality},  a case analysis is unavoidable when
interpreting our characterisation of maximality in terms of concrete relation: our theorem has the
interpretation that a concrete relation is maximal with respect to the \textsf{thins} ordering iff it is empty or it is
an equivalence relation.  But we want to avoid such a case analysis in our formal calculations.  Lemma 
\ref{TopPJTop} is crucial to our doing so.   First, we need a general lemma.
\begin{Lemma}\label{not.maximal.lemma0}{\rm \ \ \ Suppose $P$ is a per.  Then  \begin{displaymath}P{\MPperdomain}\ms{3}{=}\ms{3}(P{\MPcomp}{\MPplattop}{\MPcomp}P){\MPperdomain}~~.\end{displaymath}(Note that, in general,   $P{\MPcomp}{\MPplattop}{\MPcomp}P$ is a per if $P$ is a per; the easy proof is left to the reader.)
}%
\end{Lemma}%
{\bf Proof}~~~ 
\begin{mpdisplay}{0.15em}{6.5mm}{0mm}{2}
	$P{\MPperdomain}$\push\-\\
	$=$	\>	\>$\{$	\>\+\+\+$P$ is a per, so $P\ms{1}{=}\ms{1}P{\MPcomp}P$ (applied twice)\-\-$~~~ \}$\pop\\
	$(P{\MPcomp}P{\MPcomp}P){\MPperdomain}$\push\-\\
	$\subseteq$	\>	\>$\{$	\>\+\+\+ $P\ms{1}{\subseteq}\ms{1}{\MPplattop}$; monotonicity\-\-$~~~ \}$\pop\\
	$(P{\MPcomp}{\MPplattop}{\MPcomp}P){\MPperdomain}$\push\-\\
	$\subseteq$	\>	\>$\{$	\>\+\+\+domains (specifically $\left[\ms{2}(R{\MPcomp}S){\MPldom{}}\ms{1}{\subseteq}\ms{1}R{\MPldom{}}\ms{2}\right]$ with $R{,}S\ms{2}{:=}\ms{2}P\ms{1}{,}\ms{1}{\MPplattop}{\MPcomp}P$)\-\-$~~~ \}$\pop\\
	$P{\MPperdomain}~~.$
\end{mpdisplay}
That is, by anti-symmetry,  $P{\MPperdomain}\ms{1}{=}\ms{1}(P{\MPcomp}{\MPplattop}{\MPcomp}P){\MPperdomain}$.    
\MPendBox

We  have included the proof of lemma \ref{not.maximal.lemma0} because none of our earlier publications
document the property.  In what follows,  we frequently use well-documented 
 properties of coreflexives and domains.    Typical of the sort of properties we use is that, for all pers $P$ and
coreflexives $J$,  \begin{displaymath}{\MPplattop}{\MPcomp}P{\MPcomp}J{\MPcomp}P{\MPcomp}{\MPplattop}\ms{4}{=}\ms{4}{\MPplattop}\ms{1}{\MPcomp}\ms{1}J\ms{1}{\MPcomp}\ms{1}P{\MPperdomain}\ms{1}{\MPcomp}\ms{1}{\MPplattop}~~.\end{displaymath}The reader should be able to easily prove this property using the fact that coreflexives commute (i.e.\ for
all coreflexives $p$ and $q$,  $p{\MPcomp}q\ms{1}{=}\ms{1}q{\MPcomp}p$) and are idempotents of composition (i.e. for all coreflexives $p$, 
$p{\MPcomp}p\ms{1}{=}\ms{1}p$)   and, for all pers $P$,  $P{\MPperdomain}$ is coreflexive and 
$P{\MPperdomain}\ms{1}{\MPcomp}\ms{1}{\MPplattop}\ms{2}{=}\ms{2}P{\MPcomp}{\MPplattop}$ and ${\MPplattop}\ms{1}{\MPcomp}\ms{1}P{\MPperdomain}\ms{2}{=}\ms{2}{\MPplattop}{\MPcomp}P$.  As explained earlier, and illustrated above, we often state the
properties within the square ``everywhere'' brackets, with the convention that  $P$ ranges over pers and
lower case letters (e.g.\ $q$) range over coreflexives.

For concrete relations,  an index of a per is empty if and only if the per itself is empty.    In the absence of
the cone rule, a different idiom is needed to express such properties.  This is the function of lemma 
\ref{TopPJTop}.  
\begin{Lemma}\label{TopPJTop}{\rm \ \ \ Suppose $P$ is a per and $J$ is an index of $P$.  Then\begin{displaymath}{\MPplattop}{\MPcomp}P{\MPcomp}{\MPplattop}\ms{3}{=}\ms{3}{\MPplattop}{\MPcomp}J{\MPcomp}{\MPplattop}~~.\end{displaymath}
}%
\end{Lemma}%
{\bf Proof}~~~The proof is by mutual inclusion.
\begin{mpdisplay}{0.15em}{6.5mm}{0mm}{2}
	${\MPplattop}{\MPcomp}J{\MPcomp}{\MPplattop}$\push\-\\
	$\subseteq$	\>	\>$\{$	\>\+\+\+$J$ is an index of $P$, definition \ref{per.index}(a) and $\left[\ms{3}{\MPplattop}{\MPcomp}P\ms{3}{=}\ms{3}{\MPplattop}\ms{1}{\MPcomp}\ms{1}P{\MPperdomain}\ms{3}\right]$\-\-$~~~ \}$\pop\\
	${\MPplattop}{\MPcomp}P{\MPcomp}{\MPplattop}$\push\-\\
	$=$	\>	\>$\{$	\>\+\+\+$J$ is an index of $P$, definition \ref{per.index}(c)\-\-$~~~ \}$\pop\\
	${\MPplattop}{\MPcomp}P{\MPcomp}J{\MPcomp}P{\MPcomp}{\MPplattop}$\push\-\\
	$\subseteq$	\>	\>$\{$	\>\+\+\+$\left[\ms{1}U\ms{1}{\subseteq}\ms{1}{\MPplattop}\ms{1}\right]$ with $U\ms{1}{:=}\ms{1}{\MPplattop}{\MPcomp}P$ and $U\ms{1}{:=}\ms{1}P{\MPcomp}{\MPplattop}$\-\-$~~~ \}$\pop\\
	${\MPplattop}{\MPcomp}J{\MPcomp}{\MPplattop}~~.$
\end{mpdisplay}
\vspace{-7mm}
\MPendBox

\begin{Lemma}\label{PqIndex}{\rm \ \ \ Suppose $P$ is a per,  $q$ is coreflexive and  $J$ is an index of $P{\cup}q$.  
Then $J\ms{1}{\MPcomp}\ms{1}P{\MPperdomain}$ is an index of $P$ and $q\ms{2}{\subseteq}\ms{2}P{\MPperdomain}\ms{1}{\cup}\ms{1}J$. 
}%
\end{Lemma}%
{\bf Proof}~~~The lemma implicitly assumes that $P{\cup}q$ is a per.  This  is easily verified.  

Suppose $J$ is an index of $P{\cup}q$.  We verify the three defining
properties of an index,  \ref{per.index}(a), (b) and (c),  with $J{,}P\ms{3}{:=}\ms{3}J\ms{1}{\MPcomp}\ms{1}P{\MPperdomain}\ms{2}{,}\ms{2}P$  assuming these properties with
$J{,}P\ms{2}{:=}\ms{2}J\ms{1}{,}\ms{1}P{\cup}q$.   The first,  $J\ms{1}{\MPcomp}\ms{1}P{\MPperdomain}\ms{2}{\subseteq}\ms{2}P{\MPperdomain}$,   is immediate from the fact that $J$ is coreflexive.  For the second,
we have:
\begin{mpdisplay}{0.15em}{6.5mm}{0mm}{2}
	$J\ms{1}{\MPcomp}\ms{1}P{\MPperdomain}\ms{1}{\MPcomp}\ms{1}P\ms{1}{\MPcomp}\ms{1}J\ms{1}{\MPcomp}\ms{1}P{\MPperdomain}$\push\-\\
	$=$	\>	\>$\{$	\>\+\+\+$J$ and $P{\MPperdomain}$ are coreflexive\-\-$~~~ \}$\pop\\
	$J\ms{1}{\MPcomp}\ms{1}P{\MPperdomain}\ms{1}{\MPcomp}\ms{1}P\ms{1}{\MPcomp}\ms{1}P{\MPperdomain}\ms{1}{\MPcomp}\ms{1}J$\push\-\\
	$=$	\>	\>$\{$	\>\+\+\+$P{\MPperdomain}$ and $q$  are coreflexive, and $P{\MPperdomain}\ms{1}{\subseteq}\ms{1}P$, so\\
	$P{\MPperdomain}\ms{1}{\MPcomp}\ms{1}q\ms{1}{\MPcomp}\ms{1}P{\MPperdomain}\ms{4}{\subseteq}\ms{4}P{\MPperdomain}\ms{1}{\MPcomp}\ms{1}P{\MPperdomain}\ms{4}{=}\ms{4}P{\MPperdomain}\ms{1}{\MPcomp}\ms{1}P{\MPperdomain}\ms{1}{\MPcomp}\ms{1}P{\MPperdomain}\ms{4}{\subseteq}\ms{4}P{\MPperdomain}\ms{1}{\MPcomp}\ms{1}P\ms{1}{\MPcomp}\ms{1}P{\MPperdomain}$\-\-$~~~ \}$\pop\\
	$J\ms{1}{\MPcomp}\ms{1}(P{\MPperdomain}\ms{1}{\MPcomp}\ms{1}P\ms{1}{\MPcomp}\ms{1}P{\MPperdomain}\ms{3}{\cup}\ms{3}P{\MPperdomain}\ms{1}{\MPcomp}\ms{1}q\ms{1}{\MPcomp}\ms{1}P{\MPperdomain})\ms{1}{\MPcomp}\ms{1}J$\push\-\\
	$=$	\>	\>$\{$	\>\+\+\+distributivity\-\-$~~~ \}$\pop\\
	$J\ms{1}{\MPcomp}\ms{1}P{\MPperdomain}\ms{1}{\MPcomp}\ms{1}(P\ms{1}{\cup}\ms{1}q)\ms{1}{\MPcomp}\ms{1}P{\MPperdomain}\ms{1}{\MPcomp}\ms{1}J$\push\-\\
	$=$	\>	\>$\{$	\>\+\+\+$J$ and $P{\MPperdomain}$ are coreflexive\-\-$~~~ \}$\pop\\
	$P{\MPperdomain}\ms{1}{\MPcomp}\ms{1}J\ms{1}{\MPcomp}\ms{1}(P\ms{1}{\cup}\ms{1}q)\ms{1}{\MPcomp}\ms{1}J\ms{1}{\MPcomp}\ms{1}P{\MPperdomain}$\push\-\\
	$=$	\>	\>$\{$	\>\+\+\+$J$ is an index of $P\ms{1}{\cup}\ms{1}q$:  definition \ref{per.index}(b) with $P{,}J\ms{2}{:=}\ms{2}P{\cup}q\ms{1}{,}\ms{1}J$\-\-$~~~ \}$\pop\\
	$P{\MPperdomain}\ms{1}{\MPcomp}\ms{1}J\ms{1}{\MPcomp}\ms{1}P{\MPperdomain}$\push\-\\
	$=$	\>	\>$\{$	\>\+\+\+$J$ and $P{\MPperdomain}$ are coreflexive\-\-$~~~ \}$\pop\\
	$J\ms{1}{\MPcomp}\ms{1}P{\MPperdomain}~~.$
\end{mpdisplay}
Third,
\begin{mpdisplay}{0.15em}{6.5mm}{0mm}{2}
	$\mathsf{true}$\push\-\\
	$=$	\>	\>$\{$	\>\+\+\+$J$ is an index of $P\ms{1}{\cup}\ms{1}q$:  definition  \ref{per.index}(c) with $P{,}J\ms{2}{:=}\ms{2}P{\cup}q\ms{1}{,}\ms{1}J$ \-\-$~~~ \}$\pop\\
	$P{\cup}q\ms{3}{=}\ms{3}(P{\cup}q)\ms{1}{\MPcomp}\ms{1}J\ms{1}{\MPcomp}\ms{1}(P{\cup}q)$\push\-\\
	$\Rightarrow$	\>	\>$\{$	\>\+\+\+Leibniz\-\-$~~~ \}$\pop\\
	$P\ms{1}{\MPcomp}\ms{1}(P{\cup}q)\ms{1}{\MPcomp}\ms{1}P\ms{5}{=}\ms{5}P\ms{1}{\MPcomp}\ms{1}(P{\cup}q)\ms{1}{\MPcomp}\ms{1}J\ms{1}{\MPcomp}\ms{1}(P{\cup}q)\ms{1}{\MPcomp}\ms{1}P$\push\-\\
	$=$	\>	\>$\{$	\>\+\+\+distributivity,  $P{\MPcomp}P\ms{1}{=}\ms{1}P$\-\-$~~~ \}$\pop\\
	$P\ms{2}{\cup}\ms{2}P{\MPcomp}q{\MPcomp}P\ms{5}{=}\ms{5}(P\ms{1}{\cup}\ms{1}P{\MPcomp}q)\ms{1}{\MPcomp}\ms{1}J\ms{1}{\MPcomp}\ms{1}(P\ms{1}{\cup}\ms{1}q{\MPcomp}P)$\push\-\\
	$=$	\>	\>$\{$	\>\+\+\+ $q\ms{1}{\subseteq}\ms{1}\mathbb{I}$ and $P{\MPcomp}P\ms{1}{=}\ms{1}P$,  so  $P{\MPcomp}q{\MPcomp}P\ms{1}{\subseteq}\ms{1}P$,  $P{\MPcomp}q\ms{1}{\subseteq}\ms{1}P$  and  $q{\MPcomp}P\ms{1}{\subseteq}\ms{1}P$  \-\-$~~~ \}$\pop\\
	$P\ms{2}{=}\ms{2}P{\MPcomp}J{\MPcomp}P$\push\-\\
	$=$	\>	\>$\{$	\>\+\+\+$P{\MPperdomain}\ms{1}{\MPcomp}\ms{1}P\ms{2}{=}\ms{2}P$\-\-$~~~ \}$\pop\\
	$P\ms{4}{=}\ms{4}P\ms{1}{\MPcomp}\ms{1}J\ms{1}{\MPcomp}\ms{1}P{\MPperdomain}\ms{1}{\MPcomp}\ms{1}P~~.$
\end{mpdisplay}
This completes the proof of  the claim that  $J\ms{1}{\MPcomp}\ms{1}P{\MPperdomain}$ is an index of $P$.  Finally, 
\begin{mpdisplay}{0.15em}{6.5mm}{0mm}{2}
	$q$\push\-\\
	$\subseteq$	\>	\>$\{$	\>\+\+\+$q$ is coreflexive, so $q\ms{1}{=}\ms{1}q{\MPperdomain}$; monotonicity\-\-$~~~ \}$\pop\\
	$(P\ms{1}{\cup}\ms{1}q){\MPperdomain}$\push\-\\
	$=$	\>	\>$\{$	\>\+\+\+assumption:  $J$ is an index of $P\ms{1}{\cup}\ms{1}q$, definition \ref{per.index}(c)\-\-$~~~ \}$\pop\\
	$((P\ms{1}{\cup}\ms{1}q)\ms{1}{\MPcomp}\ms{1}J\ms{1}{\MPcomp}\ms{1}(P\ms{1}{\cup}\ms{1}q)){\MPperdomain}$\push\-\\
	$=$	\>	\>$\{$	\>\+\+\+distributivity\-\-$~~~ \}$\pop\\
	$(P{\MPcomp}J{\MPcomp}P\ms{3}{\cup}\ms{3}P{\MPcomp}J{\MPcomp}q\ms{3}{\cup}\ms{3}q{\MPcomp}J{\MPcomp}P\ms{3}{\cup}\ms{3}q{\MPcomp}J{\MPcomp}q){\MPperdomain}$\push\-\\
	$\subseteq$	\>	\>$\{$	\>\+\+\+$J\ms{1}{\subseteq}\ms{1}\mathbb{I}$ and $q\ms{1}{\subseteq}\ms{1}\mathbb{I}$,  $P{\MPcomp}P\ms{1}{\subseteq}\ms{1}P$ and monotonicity\-\-$~~~ \}$\pop\\
	$(P\ms{1}{\cup}\ms{1}J){\MPperdomain}$\push\-\\
	$=$	\>	\>$\{$	\>\+\+\+distributivity; $J$ is coreflexive,  so  $J{\MPperdomain}\ms{1}{=}\ms{1}J$\-\-$~~~ \}$\pop\\
	$P{\MPperdomain}\ms{1}{\cup}\ms{1}J~~.$
\end{mpdisplay}
\vspace{-7mm}
\MPendBox

The remaining lemmas in this section are not needed elsewhere; they are included in order to give
further insight into the nature of the \textsf{thins} ordering on pers.

Central to the notion of the \textsf{thins}  relation is that an index of a per  is found by successively ``thinning''
the relation.  More precisely,  an index  of a per is a ``thinning'' of the per  and being
an index of a per is invariant under the process of ``thinning'' the relation.  The first of these 
 two properties is  lemma \ref{index.itt}; the second is formulated in  lemma \ref{index.invariant}.  
\begin{Lemma}\label{index.invariant}{\rm \ \ \  For all pers $P$ and $Q$,  and coreflexive  $J$,  \begin{displaymath}J\mbox{ is an index of }Q\ms{6}{\Leftarrow}\ms{6}J\mbox{ is an index of }P\ms{3}{\wedge}\ms{3}P\ms{1}{\preceq}\ms{1}Q~~.\end{displaymath}
}%
\end{Lemma}%
{\bf Proof}~~~We apply theorem \ref{corefl.and.thins}:
\begin{mpdisplay}{0.15em}{6.5mm}{0mm}{2}
	$J$ is an index of $Q$\push\-\\
	$=$	\>	\>$\{$	\>\+\+\+theorem \ref{corefl.and.thins} with $P{,}Q\ms{1}{:=}\ms{1}J{,}Q$\-\-$~~~ \}$\pop\\
	$J\ms{1}{\subseteq}\ms{1}\mathbb{I}\ms{4}{\wedge}\ms{4}J\ms{1}{\preceq}\ms{1}Q$\push\-\\
	$\Leftarrow$	\>	\>$\{$	\>\+\+\+theorem \ref{itt.order} (in particular ${\preceq}$ is transitive) \-\-$~~~ \}$\pop\\
	$J\ms{1}{\subseteq}\ms{1}\mathbb{I}\ms{4}{\wedge}\ms{4}J\ms{1}{\preceq}\ms{1}P\ms{4}{\wedge}\ms{4}P\ms{1}{\preceq}\ms{1}Q$\push\-\\
	$=$	\>	\>$\{$	\>\+\+\+ theorem \ref{corefl.and.thins} with $P{,}Q\ms{1}{:=}\ms{1}J{,}P$\-\-$~~~ \}$\pop\\
	$J$ is an index of $P\ms{4}{\wedge}\ms{4}P\ms{1}{\preceq}\ms{1}Q~~.$
\end{mpdisplay}
\vspace{-7mm}
\MPendBox

Our earlier  informal interpretation of the first conjunct in the definition of \textsf{thins} is reinforced by the following
simple lemma.  Specifically,  if $J$ is an index of $P$,  $J{\MPcomp}P$ is the functional that maps a point $a$  of $P$ to the point
in $J$ that represents the equivalence class containing $a$.   If $P\ms{1}{\preceq}\ms{1}Q$ then, by lemma \ref{index.invariant},  $J$ is an
index of $Q$.  So, $J{\MPcomp}Q$ is the functional that maps a point $b$   of $Q$ to the point
in $J$ that represents the equivalence class containing $b$.  The lemma states that the two functionals agree
on points common to both $P$ and $Q$.  
\begin{Lemma}\label{common.char}{\rm \ \ \ Suppose $J$ is an index of  per $P$.   Then, for all pers $Q$,\begin{displaymath}J{\MPcomp}P\ms{3}{=}\ms{3}J\ms{1}{\MPcomp}\ms{1}Q\ms{1}{\MPcomp}\ms{1}P{\MPperdomain}\ms{8}{\Leftarrow}\ms{8}P\ms{1}{\preceq}\ms{1}Q~~.\end{displaymath}
}%
\end{Lemma}%
{\bf Proof}~~~
\begin{mpdisplay}{0.15em}{6.5mm}{0mm}{2}
	$J{\MPcomp}P$\push\-\\
	$=$	\>	\>$\{$	\>\+\+\+assumption:  $P\ms{1}{\preceq}\ms{1}Q$, definition \ref{itt.def}\-\-$~~~ \}$\pop\\
	$J\ms{1}{\MPcomp}\ms{1}P{\MPperdomain}\ms{1}{\MPcomp}\ms{1}Q\ms{1}{\MPcomp}\ms{1}P{\MPperdomain}$\push\-\\
	$=$	\>	\>$\{$	\>\+\+\+$J$ is an index of $P$,  so,  by  definition \ref{per.index}(a),  $J\ms{1}{\MPcomp}\ms{1}P{\MPperdomain}\ms{2}{=}\ms{2}J$\-\-$~~~ \}$\pop\\
	$J\ms{1}{\MPcomp}\ms{1}Q\ms{1}{\MPcomp}\ms{1}P{\MPperdomain}~~.$
\end{mpdisplay}
\vspace{-7mm}
\MPendBox

The final lemma in this section gives further insight into the relation between the \textsf{thins} relation and
indexes.
\begin{Lemma}\label{common.index}{\rm \ \ \ Suppose $P$ and $Q$ have a common index and $P\ms{1}{\subseteq}\ms{1}Q$.  Then $P\ms{1}{\preceq}\ms{1}Q$.
}%
\end{Lemma}%
{\bf Proof}~~~Suppose $J$ is an index of both $P$ and $Q$, and $P\ms{1}{\subseteq}\ms{1}Q$.  The definition of the thins relation demands
that we prove two properties.  First,
\begin{mpdisplay}{0.15em}{6.5mm}{0mm}{2}
	$Q$\push\-\\
	$=$	\>	\>$\{$	\>\+\+\+$Q$ is a per,   $\left[\ms{2}P\ms{1}{=}\ms{1}P{\MPcomp}P\ms{2}\right]$ with $P\ms{1}{:=}\ms{1}Q$\-\-$~~~ \}$\pop\\
	$Q{\MPcomp}Q{\MPcomp}Q$\push\-\\
	$\supseteq$	\>	\>$\{$	\>\+\+\+assumption:  $P\ms{1}{\subseteq}\ms{1}Q$, monotonicity\-\-$~~~ \}$\pop\\
	$Q{\MPcomp}P{\MPcomp}Q$\push\-\\
	$\supseteq$	\>	\>$\{$	\>\+\+\+$P$ is a per, so $P{\MPperdomain}\ms{1}{\subseteq}\ms{1}P$\-\-$~~~ \}$\pop\\
	$Q\ms{1}{\MPcomp}\ms{1}P{\MPperdomain}\ms{1}{\MPcomp}\ms{1}Q$\push\-\\
	$\supseteq$	\>	\>$\{$	\>\+\+\+assumption:  $J$ is an index of $P$,  definition \ref{per.index}(a)\-\-$~~~ \}$\pop\\
	$Q{\MPcomp}J{\MPcomp}Q$\push\-\\
	$=$	\>	\>$\{$	\>\+\+\+assumption:  $J$ is an index of $Q$ ,  definition \ref{per.index}(c) (with $P\ms{1}{:=}\ms{1}Q$)\-\-$~~~ \}$\pop\\
	$Q~~.$
\end{mpdisplay}
We conclude, by anti-symmetry, that $Q\ms{2}{=}\ms{2}Q\ms{1}{\MPcomp}\ms{1}P{\MPperdomain}\ms{1}{\MPcomp}\ms{1}Q$.  Now for the second property,
\begin{mpdisplay}{0.15em}{6.5mm}{0mm}{2}
	$P{\MPperdomain}\ms{1}{\MPcomp}\ms{1}Q\ms{1}{\MPcomp}\ms{1}P{\MPperdomain}$\push\-\\
	$\supseteq$	\>	\>$\{$	\>\+\+\+assumption:  $P\ms{1}{\subseteq}\ms{1}Q$, monotonicity\-\-$~~~ \}$\pop\\
	$P{\MPperdomain}\ms{1}{\MPcomp}\ms{1}P\ms{1}{\MPcomp}\ms{1}P{\MPperdomain}$\push\-\\
	$=$	\>	\>$\{$	\>\+\+\+domains\-\-$~~~ \}$\pop\\
	$P$\push\-\\
	$=$	\>	\>$\{$	\>\+\+\+assumption:  $J$ is an index of $P$,  definition \ref{per.index}(c)\-\-$~~~ \}$\pop\\
	$P{\MPcomp}J{\MPcomp}P$\push\-\\
	$=$	\>	\>$\{$	\>\+\+\+assumption:  $J$ is an index of $Q$ ,  definition \ref{per.index}(b) (with $P\ms{1}{:=}\ms{1}Q$)\-\-$~~~ \}$\pop\\
	$P{\MPcomp}J{\MPcomp}Q{\MPcomp}J{\MPcomp}P$\push\-\\
	$=$	\>	\>$\{$	\>\+\+\+$Q$ is a per,   $\left[\ms{2}P\ms{1}{=}\ms{1}P{\MPcomp}P\ms{2}\right]$ with $P\ms{1}{:=}\ms{1}Q$\-\-$~~~ \}$\pop\\
	$P{\MPcomp}J{\MPcomp}Q{\MPcomp}Q{\MPcomp}Q{\MPcomp}J{\MPcomp}P$\push\-\\
	$\supseteq$	\>	\>$\{$	\>\+\+\+assumption:  $P\ms{1}{\subseteq}\ms{1}Q$, monotonicity\-\-$~~~ \}$\pop\\
	$P{\MPcomp}J{\MPcomp}P{\MPcomp}Q{\MPcomp}P{\MPcomp}J{\MPcomp}P$\push\-\\
	$=$	\>	\>$\{$	\>\+\+\+assumption:  $J$ is an index of $P$,  definition \ref{per.index}(c)\-\-$~~~ \}$\pop\\
	$P{\MPcomp}Q{\MPcomp}P$\push\-\\
	$\supseteq$	\>	\>$\{$	\>\+\+\+$P$ is a per,  $\left[\ms{2}P{\MPperdomain}\ms{1}{\subseteq}\ms{1}P\ms{2}\right]$\-\-$~~~ \}$\pop\\
	$P{\MPperdomain}\ms{1}{\MPcomp}\ms{1}Q\ms{1}{\MPcomp}\ms{1}P{\MPperdomain}~~.$
\end{mpdisplay}
We conclude by anti-symmetry that $P{\MPperdomain}\ms{1}{\MPcomp}\ms{1}Q\ms{1}{\MPcomp}\ms{1}P{\MPperdomain}\ms{2}{=}\ms{2}P$.    Combining the two calculations, we have shown
that $P\ms{1}{\preceq}\ms{1}Q$.  (See definition \ref{itt.def}.)
\MPendBox

\section{Minimal Pers}\label{Itt:Minimal and Maximal Pers}

Our goal in this section is to characterise  the pers that are minimal with respect to the 
\textsf{thins} ordering on pers.  Section \ref{Itt:Maximality} is about
characterising   the pers that are maximal.   The notions of minimality and maximality with respect to
an ordering relation are well known.   For completeness the definition is given below.  
\begin{Definition}\label{minimal.maximal.def}{\rm \ \ \ Suppose ${\sqsubseteq}$ is a partial ordering on some set $X$.    With $x$ and $y$
ranging over elements of $X$,  we say that $y$  is \emph{minimal} with respect to the ordering
iff \begin{displaymath}{\left\langle\forall{}x\ms{3}{:}\ms{3}x\ms{1}{\sqsubseteq}\ms{1}y\ms{3}{:}\ms{3}x\ms{1}{=}\ms{1}y\right\rangle}\end{displaymath}and we say that $x$  is \emph{maximal} with respect to the ordering
iff \begin{displaymath}{\left\langle\forall{}y\ms{3}{:}\ms{3}x\ms{1}{\sqsubseteq}\ms{1}y\ms{3}{:}\ms{3}x\ms{1}{=}\ms{1}y\right\rangle}~~.\end{displaymath}\vspace{-7mm}
}
\MPendBox\end{Definition}

We  apply  definition \ref{minimal.maximal.def} in this section and in section  \ref{Itt:Maximality} to the 
\textsf{thins} ordering on pers; in section \ref{General.thins.relation} we apply the definition to the 
(yet-to-be-introduced) \textsf{thins} ordering on arbitrary relations.

A straightforward observation is that coreflexives are minimal:

\begin{Lemma}\label{corefl.is.minimal}{\rm \ \ \ \begin{displaymath}{\left\langle\forall{}Q\ms{4}{:}{:}\ms{4}\mathsf{minimal}{.}Q\ms{2}{\Leftarrow}\ms{2}Q\ms{1}{\subseteq}\ms{1}\mathbb{I}\right\rangle}~~.\end{displaymath}
}%
\end{Lemma}%
{\bf Proof}~~~Suppose $Q\ms{1}{\subseteq}\ms{1}\mathbb{I}$ and  $P\ms{1}{\preceq}\ms{1}Q$.  We prove that $P\ms{1}{=}\ms{1}Q$.  
\begin{mpdisplay}{0.15em}{6.5mm}{0mm}{2}
	$P\ms{1}{=}\ms{1}Q$\push\-\\
	$=$	\>	\>$\{$	\>\+\+\+anti-symmetry\-\-$~~~ \}$\pop\\
	$P\ms{1}{\subseteq}\ms{1}Q\ms{3}{\wedge}\ms{3}Q\ms{1}{\subseteq}\ms{1}P$\push\-\\
	$=$	\>	\>$\{$	\>\+\+\+assumption: $P\ms{1}{\preceq}\ms{1}Q$~; hence, by (\ref{itt.imp.atmost}),  $P\ms{1}{\subseteq}\ms{1}Q$\-\-$~~~ \}$\pop\\
	$Q\ms{1}{\subseteq}\ms{1}P$\push\-\\
	$=$	\>	\>$\{$	\>\+\+\+assumption: $P\ms{1}{\preceq}\ms{1}Q$~; hence by definition \ref{itt.def},  $Q\ms{2}{=}\ms{2}Q\ms{1}{\MPcomp}\ms{1}P{\MPperdomain}\ms{1}{\MPcomp}\ms{1}Q$\-\-$~~~ \}$\pop\\
	$Q\ms{1}{\MPcomp}\ms{1}P{\MPperdomain}\ms{1}{\MPcomp}\ms{1}Q\ms{2}{\subseteq}\ms{2}P$\push\-\\
	$\Leftarrow$	\>	\>$\{$	\>\+\+\+assumption:  $Q\ms{1}{\subseteq}\ms{1}\mathbb{I}$~; monotonicity  \-\-$~~~ \}$\pop\\
	$P{\MPperdomain}\ms{1}{\subseteq}\ms{1}P$\push\-\\
	$=$	\>	\>$\{$	\>\+\+\+$P$ is a per\-\-$~~~ \}$\pop\\
	$\mathsf{true}~~.$
\end{mpdisplay}
\vspace{-9mm}
\MPendBox

Lemma \ref{corefl.is.minimal} suggests that we explore the circumstances in which all minimal elements are
coreflexives.  We show that this is equivalent to the axiom of choice introduced in \cite{VB2022,VB2023a}.
\begin{Theorem}\label{choice.defs.minimal}{\rm \ \ \ The axiom of choice, axiom \ref{Axiom of Choice},   is equivalent to \begin{equation}\label{minimal.ass}
{\left\langle\forall{}P\ms{2}{:}{:}\ms{2}\mathsf{minimal}{.}P\ms{3}{\equiv}\ms{3}P\ms{1}{\subseteq}\ms{1}\mathbb{I}\right\rangle}\ms{5}{\wedge}\ms{5}{\left\langle\forall{}Q\ms{3}{:}{:}\ms{3}{\left\langle\exists{}P\ms{2}{:}\ms{2}\mathsf{minimal}{.}P\ms{2}{:}\ms{2}P\ms{1}{\preceq}\ms{1}Q\right\rangle}\right\rangle}
\end{equation}
}
\end{Theorem}
{\bf Proof}~~~The proof is by mutual implication.  First, assume (\ref{minimal.ass}).  Suppose $Q$  is
an arbitrary per.  We prove that $Q$ has an   index.  

By assumption,  there exists a per $P$ such that $P\ms{1}{\subseteq}\ms{1}\mathbb{I}$ and $P\ms{1}{\preceq}\ms{1}Q$.  Theorem  \ref{corefl.and.thins}
proves that $P$ is an   index of $Q$.  

Now assume the axiom of choice.  We must prove  (\ref{minimal.ass}).   We begin with the property \begin{equation}\label{choice.defs.minimal0}
{\left\langle\forall{}P\ms{2}{:}{:}\ms{2}\mathsf{minimal}{.}P\ms{3}{\equiv}\ms{3}P\ms{1}{\subseteq}\ms{1}\mathbb{I}\right\rangle}~~.
\end{equation}By lemma \ref{corefl.is.minimal}, it suffices to prove the implication.  Suppose $P$ is minimal.  
By the axiom of choice, $P$ has an index,  $J$ say.  Then
\begin{mpdisplay}{0.15em}{6.5mm}{0mm}{2}
	$\mathsf{true}$\push\-\\
	$=$	\>	\>$\{$	\>\+\+\+$J$ is an index of $P$;  lemma \ref{index.itt}\-\-$~~~ \}$\pop\\
	$J\ms{1}{\preceq}\ms{1}P$\push\-\\
	$\Rightarrow$	\>	\>$\{$	\>\+\+\+$P$ is minimal; definition \ref{minimal.maximal.def} \-\-$~~~ \}$\pop\\
	$P\ms{1}{=}\ms{1}J$\push\-\\
	$\Rightarrow$	\>	\>$\{$	\>\+\+\+$J$ is an index of $P$,  so,  by  definition \ref{per.index}(a),   $J\ms{1}{\subseteq}\ms{1}P{\MPperdomain}$;    $\left[\ms{2}P{\MPperdomain}\ms{1}{\subseteq}\ms{1}\mathbb{I}\ms{2}\right]$\-\-$~~~ \}$\pop\\
	$P\ms{1}{\subseteq}\ms{1}\mathbb{I}~~.$
\end{mpdisplay}
We have thus proved (\ref{choice.defs.minimal0}).  Now we consider the property\begin{equation}\label{choice.defs.minimal1}
{\left\langle\forall{}Q\ms{3}{:}{:}\ms{3}{\left\langle\exists{}P\ms{2}{:}\ms{2}\mathsf{minimal}{.}P\ms{2}{:}\ms{2}P\ms{1}{\preceq}\ms{1}Q\right\rangle}\right\rangle}~~.
\end{equation}This is established by choosing, for given $Q$,  an index of $Q$.  Indeed, if $P$  is an index of $Q$,  then, by lemma 
\ref{index.itt}, $P\ms{1}{\preceq}\ms{1}Q$ and, by (\ref{choice.defs.minimal0}), it is minimal.
\MPendBox

\begin{Example}\label{counterexample.choice.defs.minimal}{\rm \ \ \ As observed in example 
\ref{counterexample.axiom.of.choice},  the algebras of example \ref{simplest.point-free.algebras} with at least
three elements do not satisfy our axiom of choice.   Consequently,  they do not satisfy the minimality
property of theorem \ref{choice.defs.minimal}:  in both cases, ${\MPplattop}$ is minimal but not coreflexive. (See example
\ref{example.thins}.)
}%
\MPendBox\end{Example}

\section{Maximal Pers}\label{Itt:Maximality}

In this section we formulate a necessary and sufficient condition guaranteeing that a given per is
maximal with respect to the \textsf{thins} ordering.  See theorem \ref{maximal.char}.

It is relatively straightforward to show that ${\MPplatbottom}$ is maximal and all equivalence relations are maximal. 
This suggests the conjecture that these are the only maximal elements.  
Rather than formulate a proof that involves a case analysis,  we prove a more general property. 
 Specifically, we prove that a per $P$ is maximal iff $\mathbb{I}\ms{1}{\cap}\ms{1}{\MPplattop}{\MPcomp}P{\MPcomp}{\MPplattop}\ms{2}{\subseteq}\ms{2}P$.  If the cone rule
holds, $\mathbb{I}\ms{1}{\cap}\ms{1}{\MPplattop}{\MPcomp}P{\MPcomp}{\MPplattop}\ms{2}{\subseteq}\ms{2}P$ is equivalent to $\mathbb{I}\ms{1}{\subseteq}\ms{1}P\ms{2}{\vee}\ms{2}P\ms{1}{=}\ms{1}{\MPplatbottom}$.  The additional generality comes from instances of
relation algebra where the cone rule does not hold.

\subsection{Sufficient Condition for Maximality}\label{SufficientConditionforMaximality}

First, we prove the ``if'' statement.

\begin{Lemma}\label{Jules.if}{\rm \ \ \ Per $P$ is maximal if $\mathbb{I}\ms{1}{\cap}\ms{1}{\MPplattop}{\MPcomp}P{\MPcomp}{\MPplattop}\ms{2}{\subseteq}\ms{2}P$.
}%
\end{Lemma}%
{\bf Proof}~~~Suppose $\mathbb{I}\ms{1}{\cap}\ms{1}{\MPplattop}{\MPcomp}P{\MPcomp}{\MPplattop}\ms{2}{\subseteq}\ms{2}P$ and $P\ms{1}{\preceq}\ms{1}Q$.  To show that $P$ is maximal we must show that $P\ms{1}{=}\ms{1}Q$.
We first show that $P{\MPperdomain}\ms{1}{=}\ms{1}Q{\MPperdomain}$.  
\begin{mpdisplay}{0.15em}{6.5mm}{0mm}{2}
	$Q{\MPperdomain}$\push\-\\
	$=$	\>	\>$\{$	\>\+\+\+$Q$ is a per\-\-$~~~ \}$\pop\\
	$\mathbb{I}\ms{1}{\cap}\ms{1}Q$\push\-\\
	$=$	\>	\>$\{$	\>\+\+\+$P\ms{1}{\preceq}\ms{1}Q$, lemma \ref{itt.ABA}\-\-$~~~ \}$\pop\\
	$\mathbb{I}\ms{2}{\cap}\ms{2}Q{\MPcomp}P{\MPcomp}Q$\push\-\\
	$\subseteq$	\>	\>$\{$	\>\+\+\+$Q\ms{1}{\subseteq}\ms{1}{\MPplattop}$ and monotonicity\-\-$~~~ \}$\pop\\
	$\mathbb{I}\ms{2}{\cap}\ms{2}{\MPplattop}{\MPcomp}P{\MPcomp}{\MPplattop}$\push\-\\
	$=$	\>	\>$\{$	\>\+\+\+domains\-\-$~~~ \}$\pop\\
	$(\mathbb{I}\ms{2}{\cap}\ms{2}{\MPplattop}{\MPcomp}P{\MPcomp}{\MPplattop}){\MPperdomain}$\push\-\\
	$\subseteq$	\>	\>$\{$	\>\+\+\+assumption: $\mathbb{I}\ms{1}{\cap}\ms{1}{\MPplattop}{\MPcomp}P{\MPcomp}{\MPplattop}\ms{2}{\subseteq}\ms{2}P$ and monotonicity\-\-$~~~ \}$\pop\\
	$P{\MPperdomain}$\push\-\\
	$\subseteq$	\>	\>$\{$	\>\+\+\+assumption: $P\ms{1}{\preceq}\ms{1}Q$,  (\ref{itt.imp.atmost}) and monotonicity\-\-$~~~ \}$\pop\\
	$Q{\MPperdomain}$
\end{mpdisplay}
Thus $P{\MPperdomain}\ms{1}{=}\ms{1}Q{\MPperdomain}$ follows by anti-symmetry.  Now we show that $P\ms{1}{=}\ms{1}Q$:
\begin{mpdisplay}{0.15em}{6.5mm}{0mm}{2}
	$P$\push\-\\
	$=$	\>	\>$\{$	\>\+\+\+assumption: $P\ms{1}{\preceq}\ms{1}Q$\-\-$~~~ \}$\pop\\
	$P{\MPperdomain}\ms{1}{\MPcomp}\ms{1}Q\ms{1}{\MPcomp}\ms{1}P{\MPperdomain}$\push\-\\
	$=$	\>	\>$\{$	\>\+\+\+$P{\MPperdomain}\ms{1}{=}\ms{1}Q{\MPperdomain}$ (just proved) \-\-$~~~ \}$\pop\\
	$Q{\MPperdomain}\ms{1}{\MPcomp}\ms{1}Q\ms{1}{\MPcomp}\ms{1}Q{\MPperdomain}$\push\-\\
	$=$	\>	\>$\{$	\>\+\+\+domains\-\-$~~~ \}$\pop\\
	$Q~~.$
\end{mpdisplay}
\vspace{-7mm}
\MPendBox

\subsection{Necessary Condition for Maximality}\label{NecessaryConditionforMaximality}

We now turn to the converse of lemma \ref{Jules.if}.   The key fact  is theorem  \ref{not.maximal}.

\begin{Theorem}\label{not.maximal}{\rm \ \ \ Suppose $P$ is an arbitrary per.  Then, assuming the axiom of 
choice,  there is a per  $Q$ such that $\mathbb{I}\ms{1}{\cap}\ms{1}{\MPplattop}{\MPcomp}Q{\MPcomp}{\MPplattop}\ms{2}{\subseteq}\ms{2}Q$ and  $P\ms{1}{\preceq}\ms{1}Q$ .
}
\MPendBox\end{Theorem}

The remainder of this section is about proving this theorem.  Throughout 
 we assume that $P$ is an arbitrary per and $q$, $J$,  $R$   and $Q$ are defined 
by  (\ref{def.q}), (\ref{def.J}),   (\ref{def.R}) and  (\ref{Q.newdef}):\begin{equation}\label{def.q}
q\ms{4}{=}\ms{4}\mathbb{I}\ms{2}{\cap}\ms{2}{\MPplattop}{\MPcomp}P{\MPcomp}{\MPplattop}\mbox{~~,}
\end{equation}
\vspace{-5mm}
\begin{equation}\label{def.J}
J\mbox{ is an index of }P{\MPcomp}{\MPplattop}{\MPcomp}P\ms{2}{\cup}\ms{2}q\mbox{~~,}
\end{equation}
\vspace{-5mm}
\begin{equation}\label{def.R}
R\ms{3}{=}\ms{3}J{\MPcomp}{\MPplattop}{\MPcomp}J\mbox{~~,}
\end{equation}
\vspace{-5mm}
\begin{equation}\label{Q.newdef}
Q\ms{5}{=}\ms{5}P\ms{2}{\cup}\ms{2}R\ms{2}{\cup}\ms{2}P{\MPcomp}R\ms{2}{\cup}\ms{2}R{\MPcomp}P~~.
\end{equation}

The heuristics  that lead to these definitions are as follows. These
heuristics are based on the interpretation of the elements of our algebra as concrete relations and, as
such, contain implicitly several properties on which we do not rely.  Great care must therefore be taken
with such interpretations. 

Suppose $P$ is an arbitrary  per.  To prove theorem \ref{not.maximal} we have to construct a per $Q$ with two
properties, the first  of which is $\mathbb{I}\ms{1}{\cap}\ms{1}{\MPplattop}{\MPcomp}Q{\MPcomp}{\MPplattop}\ms{2}{\subseteq}\ms{2}Q$.   For concrete relations, the interpretation of this
property is that $Q$ is either the empty relation, or that $Q$ is an equivalence relation (a per with domain $\mathbb{I}$).  
For concrete relations, this leads to a case analysis on whether or not $P$ is the empty relation:  in the case
that $P$ \emph{is} the empty relation, $Q$ is  defined to also be the empty relation, and in the case that $P$ is \emph{not}
the empty relation, it is necessary to define $Q$ to be an equivalence relation.  In what follows, we assume
that $P$ is non-empty.

The second requirement on $Q$ is $P\ms{1}{\preceq}\ms{1}Q$.  In order to achieve this additional goal,  we aim  to define $Q$ to be
an equivalence relation   that extends $P$ by adding all points not in the domain of $P$ to one of the
equivalence classes of $P$.  This goal involves several elements.  Extending $P$ means defining $Q$ as $P{\cup}U$ for
some $U$; adding points to one of the equivalence classes of $P$ entails choosing one such class; finally, 
ensuring that $Q$ is an equivalence relation means guaranteeing that the domain of $Q$ is $\mathbb{I}$ and $Q$ is both
symmetric and transitive.

In general,  $P{\cup}U$ is not a per, even if $U$ is a per.  The required transitivity is readily satisfied if  $Q$ is defined 
as $(P{\cup}R)^{+}$   for some $R$; moreover the required symmetry is also satisfied if we choose for $R$ a symmetric
relation --- in particular, if $R$ is itself a per\footnote{For arbitrary (homogeneous)  relation  $U$,   $U^{+}$
denotes the transitive closure of $U$ --- the smallest relation that includes $U$ and is transitive.  For the
purposes of the current informal account,  we  assume familiarity with properties of $U^{+}$.  For example,
we assume the reader is familiar with the property that, for all $U$,  $(U^{+})^{\MPrev}\ms{1}{=}\ms{1}(U^{\MPrev})^{+}$.  The formal
calculations do not make any such assumption.}.  
Importantly, if $P$ and $R$ are both pers,    $(P{\cup}R)^{+}$ is  the smallest per that contains both $P$ and $R$.  

The goal becomes to define $Q$ to be $(P{\cup}R)^{+}$ where  $R$ is a per so defined  that its domain includes one of the
equivalence classes of $P$ together with all points that are not in the domain of $P$.  Choosing one of the
eqivalence classes of $P$ is achieved by using our axiom of choice. 

At this point, we need to anticipate the fact that we do not want to assume the cone rule.  Interpreted as
a concrete relation,  ${\MPplattop}{\MPcomp}P{\MPcomp}{\MPplattop}$ is either the empty relation or the universal relation, depending on whether
$P$ is empty or non-empty.  The  interpretation of $\mathbb{I}\ms{2}{\cap}\ms{2}{\MPplattop}{\MPcomp}P{\MPcomp}{\MPplattop}$ is thus either the empty set of points or the
set of all points.   This interpretation  relies on the cone rule, which we do not wish to exploit.   We therefore
introduce the abbreviation $q$ for  $\mathbb{I}\ms{2}{\cap}\ms{2}{\MPplattop}{\MPcomp}P{\MPcomp}{\MPplattop}$.  In the current context ---$P$ is a non-empty, concrete
relation--- $q\ms{1}{=}\ms{1}\mathbb{I}$.    References to $q$ below, rather than $\mathbb{I}$, anticipate the more general property that we
actually prove.  

Interpreted as a concrete relation,  the relation $P{\MPcomp}{\MPplattop}{\MPcomp}P$ is a per with a single equivalence class that
contains all the points in the domain of $P$.  By choosing an index  of $P{\MPcomp}{\MPplattop}{\MPcomp}P$, we effectively choose one point 
in the domain of $P$; an index $J$ of  $P{\MPcomp}{\MPplattop}{\MPcomp}P\ms{2}{\cup}\ms{2}q$ is thus interpreted as a set of points consisting of one  point in the domain of $P$ together with
all the points in $q$ that are not in the domain of $P$.  The interpretation  of $J\ms{1}{\MPcomp}\ms{1}P{\MPperdomain}$ is the point in the domain  
of $P$ chosen by $J$; the formal  properties of $J\ms{1}{\MPcomp}\ms{1}P{\MPperdomain}$ play a central role in the proof.

The interpretation of $J{\MPcomp}{\MPplattop}{\MPcomp}J$   is  a per with exactly one
equivalence class: the class containing all the points equivalent in $P$ to the point  $J\ms{1}{\MPcomp}\ms{1}P{\MPperdomain}$
together with all the  points in $q$  that are not in the domain of $P$.   Thus $J{\MPcomp}{\MPplattop}{\MPcomp}J$ is the per $R$ needed to
achieve our objective.

There is one more problem to be resolved.  We want to define $Q$ to be the transitive closure of  $P\ms{1}{\cup}\ms{1}R$: the 
least transitive relation that  includes both $P$ and $R$ (and hence, since $R$ is a per, the least per that
includes both $P$ and $R$).  But, for arbitrary concrete relation $U$,  its transitive closure $U^{+}$ is  
computed by a possibly non-terminating process:  beginning with $U$ continually add to $U$ successive
powers of $U$  (thus computing $U$,  $U\ms{1}{\cup}\ms{1}U{\MPcomp}U$,  $U\ms{1}{\cup}\ms{1}U{\MPcomp}U\ms{1}{\cup}\ms{1}U{\MPcomp}U{\MPcomp}U$, and so on).  Fortunately,  in the case of 
$P\ms{1}{\cup}\ms{1}R$ where $R$ is  defined by (\ref{def.R}), this process is terminating; indeed, it terminates after the first
iteration.  The definition of  $Q$  above anticipates this fact.  See section \ref{Qprops}. 

Note that, in the above informal account,  we have been obliged  to refer to ``points'' and to points
``not'' having a certain property.  It is important to note that our calculations make no assumptions about
the existence of points or the existence of complements.  The above informal account is applicable only to
concrete relations.

Let us now proceed with the proof of theorem \ref{not.maximal} where $Q$ is defined by (\ref{Q.newdef}).  In section 
\ref{PointChoice} we establish a number of properties of the index $J$;  that   $Q$ is a per is proved 
in section \ref{Qprops}; finally,  section \ref{ProofMainTheorem} is where theorem \ref{not.maximal} is proved.
 
\subsubsection{Properties of the Index $J$}\label{PointChoice}

The focus in this subsection is on the properties of the index $J$.  
The definition of $J$  assumes that $P{\MPcomp}{\MPplattop}{\MPcomp}P\ms{2}{\cup}\ms{2}q$ is a per.  This is  a straightforward consequence of the fact
that, in general,  $P\ms{1}{\cup}\ms{1}q$ is a per if $P$ is a per and $q$ is coreflexive,  and $P{\MPcomp}{\MPplattop}{\MPcomp}P$ is a per if $P$ is a per.  The details
are left to the reader.

We  now proceed to exploit definition (\ref{def.J}).  A key property is  lemma  \ref{JP.index};
other lemmas are used either to establish these two lemmas or in later calculations.  
\begin{Lemma}\label{q.domain}{\rm \ \ \ \begin{displaymath}(P{\MPcomp}{\MPplattop}{\MPcomp}P\ms{2}{\cup}\ms{2}q){\MPperdomain}\ms{3}{=}\ms{3}q~~.\end{displaymath}
}%
\end{Lemma}%
{\bf Proof}~~~
\begin{mpdisplay}{0.15em}{6.5mm}{0mm}{2}
	$(P{\MPcomp}{\MPplattop}{\MPcomp}P\ms{2}{\cup}\ms{2}q){\MPperdomain}$\push\-\\
	$=$	\>	\>$\{$	\>\+\+\+distributivity property of domain operator\-\-$~~~ \}$\pop\\
	$(P{\MPcomp}{\MPplattop}{\MPcomp}P){\MPperdomain}\ms{2}{\cup}\ms{2}q{\MPperdomain}$\push\-\\
	$=$	\>	\>$\{$	\>\+\+\+lemma \ref{not.maximal.lemma0} and $q$ is coreflexive\-\-$~~~ \}$\pop\\
	$P{\MPperdomain}\ms{2}{\cup}\ms{2}q$\push\-\\
	$=$	\>	\>$\{$	\>\+\+\+definition of $q$:  (\ref{def.q})\-\-$~~~ \}$\pop\\
	$P{\MPperdomain}\ms{2}{\cup}\ms{2}(\mathbb{I}\ms{2}{\cap}\ms{2}{\MPplattop}{\MPcomp}P{\MPcomp}{\MPplattop})$\push\-\\
	$=$	\>	\>$\{$	\>\+\+\+ $\left[\ms{3}P{\MPperdomain}\ms{2}{=}\ms{2}\mathbb{I}\ms{1}{\cap}\ms{1}P\ms{2}{\subseteq}\ms{2}\mathbb{I}\ms{2}{\cap}\ms{2}{\MPplattop}{\MPcomp}P{\MPcomp}{\MPplattop}\ms{3}\right]$\-\-$~~~ \}$\pop\\
	$\mathbb{I}\ms{2}{\cap}\ms{2}{\MPplattop}{\MPcomp}P{\MPcomp}{\MPplattop}$\push\-\\
	$=$	\>	\>$\{$	\>\+\+\+definition of $q$:  (\ref{def.q})\-\-$~~~ \}$\pop\\
	$q~~.$
\end{mpdisplay}
\vspace{-7mm}
\MPendBox

Now  $J$ is defined to be an index of $P{\MPcomp}{\MPplattop}{\MPcomp}P\ms{2}{\cup}\ms{2}q$.   So,   applying definition \ref{per.index}(a), 
 an immediate corollary of lemma \ref{q.domain} is  that \begin{equation}\label{J.atmost.q}
J\ms{1}{\subseteq}\ms{1}q~~.
\end{equation}
The interpretation of   $J\ms{1}{\MPcomp}\ms{1}P{\MPperdomain}$ in terms of concrete relations is the point   chosen by  the
index $J$   that indexes  $P{\MPcomp}{\MPplattop}{\MPcomp}P$;  the other points  in $J$ are the points in  $q$ that are not in $P{\MPperdomain}$.  Aspects of these 
informal interpretations that are relevant to our calculations  are expressed by lemma \ref{JP.index}.  

\begin{Lemma}\label{JP.index}{\rm \ \ \ $J\ms{1}{\MPcomp}\ms{1}P{\MPperdomain}$ is an index of $P{\MPcomp}{\MPplattop}{\MPcomp}P$.  Hence\begin{equation}\label{JP}
J\ms{1}{\MPcomp}\ms{1}P{\MPperdomain}\ms{4}{=}\ms{4}J\ms{1}{\MPcomp}\ms{1}P{\MPperdomain}\ms{1}{\MPcomp}\ms{1}{\MPplattop}\ms{1}{\MPcomp}\ms{1}P{\MPperdomain}\ms{1}{\MPcomp}\ms{1}J~~,
\end{equation}
\vspace{-7mm}\begin{equation}\label{PtopP}
P{\MPcomp}{\MPplattop}{\MPcomp}P\ms{4}{=}\ms{4}P\ms{1}{\MPcomp}\ms{1}{\MPplattop}\ms{1}{\MPcomp}\ms{1}J\ms{1}{\MPcomp}\ms{1}P{\MPperdomain}\ms{1}{\MPcomp}\ms{1}{\MPplattop}\ms{1}{\MPcomp}\ms{1}P~~,
\end{equation}
\vspace{-7mm}\begin{equation}\label{Jules3}
{\MPplattop}\ms{1}{\MPcomp}\ms{1}P{\MPperdomain}\ms{1}{\MPcomp}\ms{1}{\MPplattop}\ms{4}{=}\ms{4}{\MPplattop}\ms{1}{\MPcomp}\ms{1}J\ms{1}{\MPcomp}\ms{1}P{\MPperdomain}\ms{1}{\MPcomp}\ms{1}{\MPplattop}\mbox{~~, and}
\end{equation}
\vspace{-7mm}\begin{equation}\label{Jules(8)}
q\ms{3}{\subseteq}\ms{3}P{\MPperdomain}\ms{1}{\cup}\ms{1}J~~.
\end{equation}
}%
\end{Lemma}%
{\bf Proof}~~~
\begin{mpdisplay}{0.15em}{6.5mm}{0mm}{2}
	$\mathsf{true}$\push\-\\
	$=$	\>	\>$\{$	\>\+\+\+lemma \ref{PqIndex} with $J{,}P\ms{3}{:=}\ms{3}J\ms{2}{,}\ms{2}P{\MPcomp}{\MPplattop}{\MPcomp}P\ms{2}{\cup}\ms{2}q$\-\-$~~~ \}$\pop\\
	$J\ms{1}{\MPcomp}\ms{1}(P{\MPcomp}{\MPplattop}{\MPcomp}P){\MPperdomain}$ is an index of $P{\MPcomp}{\MPplattop}{\MPcomp}P$\push\-\\
	$=$	\>	\>$\{$	\>\+\+\+lemma \ref{not.maximal.lemma0}\-\-$~~~ \}$\pop\\
	$J\ms{1}{\MPcomp}\ms{1}P{\MPperdomain}$ is an index of $P{\MPcomp}{\MPplattop}{\MPcomp}P~~.$
\end{mpdisplay}
We now derive  properties (\ref{JP}),   (\ref{PtopP}),  (\ref{Jules3}) and (\ref{Jules(8)}).  First (\ref{JP}):
\begin{mpdisplay}{0.15em}{6.5mm}{0mm}{2}
	$J\ms{1}{\MPcomp}\ms{1}P{\MPperdomain}$\push\-\\
	$=$	\>	\>$\{$	\>\+\+\+definition \ref{per.index}(b) with $J{,}P\ms{3}{:=}\ms{3}J\ms{1}{\MPcomp}\ms{1}P{\MPperdomain}\ms{2}{,}\ms{2}P{\MPcomp}{\MPplattop}{\MPcomp}P$\-\-$~~~ \}$\pop\\
	$J\ms{1}{\MPcomp}\ms{1}P{\MPperdomain}\ms{1}{\MPcomp}\ms{1}P\ms{1}{\MPcomp}\ms{1}{\MPplattop}\ms{1}{\MPcomp}\ms{1}P\ms{1}{\MPcomp}\ms{1}J\ms{1}{\MPcomp}\ms{1}P{\MPperdomain}$\push\-\\
	$=$	\>	\>$\{$	\>\+\+\+$\left[\ms{3}P{\MPperdomain}\ms{1}{\MPcomp}\ms{1}P\ms{2}{=}\ms{2}P\ms{2}{=}\ms{2}P\ms{1}{\MPcomp}\ms{1}P{\MPperdomain}\ms{3}\right]$,  $J$ and $P{\MPperdomain}$ are coreflexive, so $J\ms{1}{\MPcomp}\ms{1}P{\MPperdomain}\ms{2}{=}\ms{2}P{\MPperdomain}\ms{1}{\MPcomp}\ms{1}J$\-\-$~~~ \}$\pop\\
	$J\ms{1}{\MPcomp}\ms{1}P{\MPperdomain}\ms{1}{\MPcomp}\ms{1}{\MPplattop}\ms{1}{\MPcomp}\ms{1}P{\MPperdomain}\ms{1}{\MPcomp}\ms{1}J~~.$
\end{mpdisplay}
Property (\ref{PtopP}) follows:
\begin{mpdisplay}{0.15em}{6.5mm}{0mm}{2}
	$P{\MPcomp}{\MPplattop}{\MPcomp}P$\push\-\\
	$=$	\>	\>$\{$	\>\+\+\+definition \ref{per.index}(c)  with $J{,}P\ms{3}{:=}\ms{3}J\ms{1}{\MPcomp}\ms{1}P{\MPperdomain}\ms{2}{,}\ms{2}P{\MPcomp}{\MPplattop}{\MPcomp}P$\-\-$~~~ \}$\pop\\
	$P\ms{1}{\MPcomp}\ms{1}{\MPplattop}\ms{1}{\MPcomp}\ms{1}P\ms{1}{\MPcomp}\ms{1}J\ms{1}{\MPcomp}\ms{1}P{\MPperdomain}\ms{1}{\MPcomp}\ms{1}P\ms{1}{\MPcomp}\ms{1}{\MPplattop}\ms{1}{\MPcomp}\ms{1}P$\push\-\\
	$=$	\>	\>$\{$	\>\+\+\+$\left[\ms{3}{\MPplattop}{\MPcomp}P\ms{2}{=}\ms{2}{\MPplattop}\ms{1}{\MPcomp}\ms{1}P{\MPperdomain}\ms{3}\right]$,  $\left[\ms{3}P{\MPperdomain}\ms{1}{\MPcomp}\ms{1}J\ms{2}{=}\ms{2}J\ms{1}{\MPcomp}\ms{1}P{\MPperdomain}\ms{3}\right]$,  $\left[\ms{3}P{\MPperdomain}\ms{1}{\MPcomp}\ms{1}P\ms{2}{=}\ms{2}P\ms{3}\right]$\-\-$~~~ \}$\pop\\
	$P\ms{1}{\MPcomp}\ms{1}{\MPplattop}\ms{1}{\MPcomp}\ms{1}J\ms{1}{\MPcomp}\ms{1}P{\MPperdomain}\ms{1}{\MPcomp}\ms{1}{\MPplattop}\ms{1}{\MPcomp}\ms{1}P~~.$
\end{mpdisplay}
Next (\ref{Jules3}):
\begin{mpdisplay}{0.15em}{6.5mm}{0mm}{2}
	${\MPplattop}\ms{1}{\MPcomp}\ms{1}J\ms{1}{\MPcomp}\ms{1}P{\MPperdomain}\ms{1}{\MPcomp}\ms{1}{\MPplattop}$\push\-\\
	$=$	\>	\>$\{$	\>\+\+\+lemma \ref{TopPJTop} with $J{,}P\ms{3}{:=}\ms{3}J\ms{1}{\MPcomp}\ms{1}P{\MPperdomain}\ms{2}{,}\ms{2}P{\MPcomp}{\MPplattop}{\MPcomp}P$\-\-$~~~ \}$\pop\\
	${\MPplattop}{\MPcomp}P{\MPcomp}{\MPplattop}{\MPcomp}P{\MPcomp}{\MPplattop}$\push\-\\
	$=$	\>	\>$\{$	\>\+\+\+$\left[\ms{3}{\MPplattop}{\MPcomp}P\ms{2}{=}\ms{2}{\MPplattop}\ms{1}{\MPcomp}\ms{1}P{\MPperdomain}\ms{3}\right]$ with $P\ms{1}{:=}\ms{1}P{\MPcomp}{\MPplattop}{\MPcomp}P$, lemma \ref{not.maximal.lemma0}\-\-$~~~ \}$\pop\\
	${\MPplattop}\ms{1}{\MPcomp}\ms{1}P{\MPperdomain}\ms{1}{\MPcomp}\ms{1}{\MPplattop}~~.$
\end{mpdisplay}
Finally (\ref{Jules(8)}):
\begin{mpdisplay}{0.15em}{6.5mm}{0mm}{2}
	$q\ms{3}{\subseteq}\ms{3}P{\MPperdomain}\ms{1}{\cup}\ms{1}J$\push\-\\
	$=$	\>	\>$\{$	\>\+\+\+lemma \ref{not.maximal.lemma0}\-\-$~~~ \}$\pop\\
	$q\ms{3}{\subseteq}\ms{3}(P{\MPcomp}{\MPplattop}{\MPcomp}P){\MPperdomain}\ms{1}{\cup}\ms{1}J$\push\-\\
	$=$	\>	\>$\{$	\>\+\+\+assumption:  $J$ is an index of $P{\MPcomp}{\MPplattop}{\MPcomp}P\ms{2}{\cup}\ms{2}q$,  lemma \ref{PqIndex}  with $P\ms{1}{:=}\ms{1}P{\MPcomp}{\MPplattop}{\MPcomp}P$\-\-$~~~ \}$\pop\\
	$\mathsf{true}~~.$
\end{mpdisplay}
\vspace{-7mm}
\MPendBox

If $p$ is coreflexive and non-empty,  the (concrete-relational)   interpretation of  $p{\MPcomp}{\MPplattop}{\MPcomp}p$  is a per with
domain $p$ that has exactly  one equivalence class;  the  interpretation of  (\ref{JP}) is thus 
that $J\ms{1}{\MPcomp}\ms{1}P{\MPperdomain}$ is a single point that is in both $J$ and the domain of $P$.  That is,  $J\ms{1}{\MPcomp}\ms{1}P{\MPperdomain}$ is the point in the domain
of $P$ chosen by the index $J$.  
Property  (\ref{Jules(8)}) is necessitated by our wish to avoid assuming the cone rule.  For concrete relations,
it is interpreted as the property that $P{\MPperdomain}\ms{1}{\cup}\ms{1}J$ is  $\mathbb{I}$  if $P$ is non-empty.

\subsubsection{ $Q$ is a per}\label{Qprops}

A basic requirement on $Q$ is that it is a per, i.e.\ symmetric and transitive.  
The transitivity of $Q$ is not obvious.  The complication is that we do not wish to exploit the cone
rule.  Instead, we have to prove several  lemmas that  are trivial if the cone rule is assumed: 
lemmas \ref{JtoJ},    \ref{edRTR} and \ref{TopRTop.is.TopPTop}.  Lemma \ref{PRP} is also needed.

\begin{Lemma}\label{JtoJ}{\rm \ \ \ \begin{displaymath}J{\MPcomp}{\MPplattop}{\MPcomp}J\ms{3}{=}\ms{3}J{\MPcomp}{\MPplattop}{\MPcomp}P{\MPcomp}{\MPplattop}{\MPcomp}J~~.\end{displaymath}
}%
\end{Lemma}%
{\bf Proof}~~~The proof is by mutual inclusion.
\begin{mpdisplay}{0.15em}{6.5mm}{0mm}{2}
	$J{\MPcomp}{\MPplattop}{\MPcomp}P{\MPcomp}{\MPplattop}{\MPcomp}J$\push\-\\
	$\subseteq$	\>	\>$\{$	\>\+\+\+$\left[\ms{1}U\ms{1}{\subseteq}\ms{1}{\MPplattop}\ms{1}\right]$ with $U\ms{1}{:=}\ms{1}{\MPplattop}{\MPcomp}P{\MPcomp}{\MPplattop}$, monotonicity\-\-$~~~ \}$\pop\\
	$J{\MPcomp}{\MPplattop}{\MPcomp}J$\push\-\\
	$=$	\>	\>$\{$	\>\+\+\+(\ref{J.atmost.q}),  $J$ and $q$ are coreflexive (so $J{\MPcomp}q\ms{1}{=}\ms{1}J$) \-\-$~~~ \}$\pop\\
	$J{\MPcomp}q{\MPcomp}{\MPplattop}{\MPcomp}J$\push\-\\
	$\subseteq$	\>	\>$\{$	\>\+\+\+definition of $q$:  (\ref{def.q}), monotonicity\-\-$~~~ \}$\pop\\
	$J{\MPcomp}{\MPplattop}{\MPcomp}P{\MPcomp}{\MPplattop}{\MPcomp}{\MPplattop}{\MPcomp}J$\push\-\\
	$\subseteq$	\>	\>$\{$	\>\+\+\+$\left[\ms{1}U\ms{1}{\subseteq}\ms{1}{\MPplattop}\ms{1}\right]$ with $U\ms{1}{:=}\ms{1}{\MPplattop}{\MPcomp}{\MPplattop}$\-\-$~~~ \}$\pop\\
	$J{\MPcomp}{\MPplattop}{\MPcomp}P{\MPcomp}{\MPplattop}{\MPcomp}J~~.$
\end{mpdisplay}
\vspace{-7mm}
\MPendBox

\begin{Lemma}\label{edRTR}{\rm \ \ \  \begin{displaymath}R{\MPcomp}P{\MPcomp}R\ms{4}{=}\ms{4}R\ms{4}{=}\ms{4}R\ms{1}{\MPcomp}\ms{1}P{\MPperdomain}\ms{1}{\MPcomp}\ms{1}R~~.\end{displaymath}
}%
\end{Lemma}%
{\bf Proof}~~~The proof is by mutual inclusion.  Recalling the definition of $R$, (\ref{def.R}), we  have:
\begin{mpdisplay}{0.15em}{6.5mm}{0mm}{2}
	$J{\MPcomp}{\MPplattop}{\MPcomp}J{\MPcomp}P{\MPcomp}J{\MPcomp}{\MPplattop}{\MPcomp}J$\push\-\\
	$\subseteq$	\>	\>$\{$	\>\+\+\+$\left[\ms{1}U\ms{1}{\subseteq}\ms{1}{\MPplattop}\ms{1}\right]$ with $U\ms{1}{:=}\ms{1}{\MPplattop}{\MPcomp}J{\MPcomp}P{\MPcomp}J{\MPcomp}{\MPplattop}$\-\-$~~~ \}$\pop\\
	$J{\MPcomp}{\MPplattop}{\MPcomp}J$\push\-\\
	$=$	\>	\>$\{$	\>\+\+\+lemma \ref{JtoJ}\-\-$~~~ \}$\pop\\
	$J{\MPcomp}{\MPplattop}{\MPcomp}P{\MPcomp}{\MPplattop}{\MPcomp}J$\push\-\\
	$=$	\>	\>$\{$	\>\+\+\+$\left[\ms{3}{\MPplattop}{\MPcomp}P\ms{2}{=}\ms{2}{\MPplattop}\ms{1}{\MPcomp}\ms{1}P{\MPperdomain}\ms{3}\right]$,   (\ref{Jules3})\-\-$~~~ \}$\pop\\
	$J\ms{1}{\MPcomp}\ms{1}{\MPplattop}\ms{1}{\MPcomp}\ms{1}J\ms{1}{\MPcomp}\ms{1}P{\MPperdomain}\ms{1}{\MPcomp}\ms{1}{\MPplattop}\ms{1}{\MPcomp}\ms{1}J$\push\-\\
	$=$	\>	\>$\{$	\>\+\+\+$J$ and $P{\MPperdomain}$ are coreflexive,  so $J\ms{1}{\MPcomp}\ms{1}P{\MPperdomain}\ms{2}{=}\ms{2}J\ms{1}{\MPcomp}\ms{1}J\ms{1}{\MPcomp}\ms{1}P{\MPperdomain}\ms{2}{=}\ms{2}J\ms{1}{\MPcomp}\ms{1}P{\MPperdomain}\ms{1}{\MPcomp}\ms{1}J$\-\-$~~~ \}$\pop\\
	$J\ms{1}{\MPcomp}\ms{1}{\MPplattop}\ms{1}{\MPcomp}\ms{1}J\ms{1}{\MPcomp}\ms{1}P{\MPperdomain}\ms{1}{\MPcomp}\ms{1}J\ms{1}{\MPcomp}\ms{1}{\MPplattop}\ms{1}{\MPcomp}\ms{1}J$\push\-\\
	$\subseteq$	\>	\>$\{$	\>\+\+\+$P{\MPperdomain}\ms{1}{\subseteq}\ms{1}P$ \-\-$~~~ \}$\pop\\
	$J{\MPcomp}{\MPplattop}{\MPcomp}J{\MPcomp}P{\MPcomp}J{\MPcomp}{\MPplattop}{\MPcomp}J$
\end{mpdisplay}
We conclude, by mutual inclusion, that\begin{displaymath}J{\MPcomp}{\MPplattop}{\MPcomp}J{\MPcomp}P{\MPcomp}J{\MPcomp}{\MPplattop}{\MPcomp}J\ms{4}{=}\ms{4}J{\MPcomp}{\MPplattop}{\MPcomp}J\ms{4}{=}\ms{4}J\ms{1}{\MPcomp}\ms{1}{\MPplattop}\ms{1}{\MPcomp}\ms{1}J\ms{1}{\MPcomp}\ms{1}P{\MPperdomain}\ms{1}{\MPcomp}\ms{1}J\ms{1}{\MPcomp}\ms{1}{\MPplattop}\ms{1}{\MPcomp}\ms{1}J~~.\end{displaymath}The lemma follows by instantiating  the definition of $R$.
\MPendBox

\begin{Lemma}\label{TopRTop.is.TopPTop}{\rm \ \ \ \begin{displaymath}{\MPplattop}{\MPcomp}P{\MPcomp}{\MPplattop}\ms{4}{=}\ms{4}{\MPplattop}{\MPcomp}Q{\MPcomp}{\MPplattop}~~.\end{displaymath}
}%
\end{Lemma}%
{\bf Proof}~~~
\begin{mpdisplay}{0.15em}{6.5mm}{0mm}{2}
	${\MPplattop}{\MPcomp}Q{\MPcomp}{\MPplattop}$\push\-\\
	$=$	\>	\>$\{$	\>\+\+\+definition:  (\ref{Q.newdef}) and distributivity\-\-$~~~ \}$\pop\\
	${\MPplattop}{\MPcomp}P{\MPcomp}{\MPplattop}\ms{4}{\cup}\ms{4}{\MPplattop}{\MPcomp}R{\MPcomp}{\MPplattop}\ms{4}{\cup}\ms{4}{\MPplattop}{\MPcomp}P{\MPcomp}R{\MPcomp}{\MPplattop}\ms{4}{\cup}\ms{4}{\MPplattop}{\MPcomp}R{\MPcomp}P{\MPcomp}{\MPplattop}$\push\-\\
	$=$	\>	\>$\{$	\>\+\+\+$\left[\ms{2}{\MPplattop}{\MPcomp}R\ms{1}{\subseteq}\ms{1}{\MPplattop}\ms{2}\right]$ and $\left[\ms{2}R{\MPcomp}{\MPplattop}\ms{1}{\subseteq}\ms{1}{\MPplattop}\ms{2}\right]$,   \\
	so $\left[\ms{3}{\MPplattop}{\MPcomp}P{\MPcomp}R{\MPcomp}{\MPplattop}\ms{3}{\cup}\ms{3}{\MPplattop}{\MPcomp}R{\MPcomp}P{\MPcomp}{\MPplattop}\ms{3}{\subseteq}\ms{3}{\MPplattop}{\MPcomp}P{\MPcomp}{\MPplattop}\ms{3}\right]$\-\-$~~~ \}$\pop\\
	${\MPplattop}{\MPcomp}P{\MPcomp}{\MPplattop}\ms{4}{\cup}\ms{4}{\MPplattop}{\MPcomp}R{\MPcomp}{\MPplattop}$\push\-\\
	$=$	\>	\>$\{$	\>\+\+\+lemma \ref{edRTR}\-\-$~~~ \}$\pop\\
	${\MPplattop}{\MPcomp}P{\MPcomp}{\MPplattop}\ms{4}{\cup}\ms{4}{\MPplattop}{\MPcomp}R{\MPcomp}P{\MPcomp}R{\MPcomp}{\MPplattop}$\push\-\\
	$=$	\>	\>$\{$	\>\+\+\+$\left[\ms{2}{\MPplattop}{\MPcomp}R\ms{1}{\subseteq}\ms{1}{\MPplattop}\ms{2}\right]$ and $\left[\ms{2}R{\MPcomp}{\MPplattop}\ms{1}{\subseteq}\ms{1}{\MPplattop}\ms{2}\right]$,   so $\left[\ms{3}{\MPplattop}{\MPcomp}R{\MPcomp}P{\MPcomp}R{\MPcomp}{\MPplattop}\ms{2}{\subseteq}\ms{2}{\MPplattop}{\MPcomp}P{\MPcomp}{\MPplattop}\ms{3}\right]$\-\-$~~~ \}$\pop\\
	${\MPplattop}{\MPcomp}P{\MPcomp}{\MPplattop}~~.$
\end{mpdisplay}
\vspace{-7mm}
\MPendBox

\begin{Lemma}\label{PRP}{\rm \ \ \ \begin{displaymath}P{\MPcomp}R{\MPcomp}P\ms{2}{\subseteq}\ms{2}P~~.\end{displaymath}
}%
\end{Lemma}%
{\bf Proof}~~~
\begin{mpdisplay}{0.15em}{6.5mm}{0mm}{2}
	$P{\MPcomp}R{\MPcomp}P$\push\-\\
	$=$	\>	\>$\{$	\>\+\+\+definition:  (\ref{def.R})\-\-$~~~ \}$\pop\\
	$P{\MPcomp}J{\MPcomp}{\MPplattop}{\MPcomp}J{\MPcomp}P$\push\-\\
	$=$	\>	\>$\{$	\>\+\+\+$\left[\ms{3}P\ms{1}{\MPcomp}\ms{1}P{\MPperdomain}\ms{2}{=}\ms{2}P\ms{2}{=}\ms{2}P{\MPperdomain}\ms{1}{\MPcomp}\ms{1}P\ms{3}\right]$\-\-$~~~ \}$\pop\\
	$P\ms{1}{\MPcomp}\ms{1}P{\MPperdomain}\ms{1}{\MPcomp}\ms{1}J\ms{1}{\MPcomp}\ms{1}{\MPplattop}\ms{1}{\MPcomp}\ms{1}J\ms{1}{\MPcomp}\ms{1}P{\MPperdomain}\ms{1}{\MPcomp}\ms{1}P$\push\-\\
	$=$	\>	\>$\{$	\>\+\+\+$J$ and $P{\MPperdomain}$ are coreflexive, so $P{\MPperdomain}\ms{1}{\MPcomp}\ms{1}J\ms{2}{=}\ms{2}J\ms{1}{\MPcomp}\ms{1}P{\MPperdomain}$,  (\ref{JP})\-\-$~~~ \}$\pop\\
	$P\ms{1}{\MPcomp}\ms{1}J\ms{1}{\MPcomp}\ms{1}P{\MPperdomain}\ms{1}{\MPcomp}\ms{1}P$\push\-\\
	$\subseteq$	\>	\>$\{$	\>\+\+\+$J$  and $P{\MPperdomain}$ are coreflexive,  and monotonicity\-\-$~~~ \}$\pop\\
	$P{\MPcomp}P$\push\-\\
	$=$	\>	\>$\{$	\>\+\+\+$P$ is a per\-\-$~~~ \}$\pop\\
	$P~~.$
\end{mpdisplay}
\vspace{-7mm}
\MPendBox

We now  show more than just transitivity of $Q$: we show that $Q$ is the transitive closure of $P\ms{1}{\cup}\ms{1}R$.
\begin{Lemma}\label{Pcupa}{\rm \ \ \ \begin{equation}\label{Pcupa0}
Q\ms{4}{=}\ms{4}(P\ms{1}{\cup}\ms{1}R)\ms{1}{\MPcomp}\ms{1}(P\ms{1}{\cup}\ms{1}R)\ms{4}{=}\ms{4}(P\ms{1}{\cup}\ms{1}R)\ms{1}{\MPcomp}\ms{1}P{\MPperdomain}\ms{1}{\MPcomp}\ms{1}(P\ms{1}{\cup}\ms{1}R)\mbox{~~, and}
\end{equation}
\vspace{-7mm}\begin{equation}\label{Pcupa1}
Q\ms{1}{\MPcomp}\ms{1}(P\ms{1}{\cup}\ms{1}R)\ms{3}{=}\ms{3}Q~~.
\end{equation}
}%
\end{Lemma}%
{\bf Proof}~~~Property (\ref{Pcupa0})  is easy to prove:
\begin{mpdisplay}{0.15em}{6.5mm}{0mm}{2}
	$(P\ms{1}{\cup}\ms{1}R)\ms{1}{\MPcomp}\ms{1}P{\MPperdomain}\ms{1}{\MPcomp}\ms{1}(P\ms{1}{\cup}\ms{1}R)$\push\-\\
	$=$	\>	\>$\{$	\>\+\+\+distributivity,  $P\ms{1}{\MPcomp}\ms{1}P{\MPperdomain}\ms{2}{=}\ms{2}P\ms{2}{=}\ms{2}P{\MPperdomain}\ms{1}{\MPcomp}\ms{1}P$ and $P\ms{1}{=}\ms{1}P{\MPcomp}P$\-\-$~~~ \}$\pop\\
	$P\ms{3}{\cup}\ms{3}R{\MPcomp}P\ms{3}{\cup}\ms{3}P{\MPcomp}R\ms{3}{\cup}\ms{3}R\ms{1}{\MPcomp}\ms{1}P{\MPperdomain}\ms{1}{\MPcomp}\ms{1}R$\push\-\\
	$=$	\>	\>$\{$	\>\+\+\+lemma \ref{edRTR}\-\-$~~~ \}$\pop\\
	$P\ms{2}{\cup}\ms{2}R{\MPcomp}P\ms{2}{\cup}\ms{2}P{\MPcomp}R\ms{2}{\cup}\ms{2}R$\push\-\\
	$=$	\>	\>$\{$	\>\+\+\+$P$ and $R$ are pers, so $P\ms{1}{=}\ms{1}P{\MPcomp}P$ and $R\ms{1}{=}\ms{1}R{\MPcomp}R$\-\-$~~~ \}$\pop\\
	$P{\MPcomp}P\ms{2}{\cup}\ms{2}R{\MPcomp}R\ms{2}{\cup}\ms{2}P{\MPcomp}R\ms{2}{\cup}\ms{2}R{\MPcomp}P$\push\-\\
	$=$	\>	\>$\{$	\>\+\+\+distributivity\-\-$~~~ \}$\pop\\
	$(P\ms{1}{\cup}\ms{1}R)\ms{1}{\MPcomp}\ms{1}(P\ms{1}{\cup}\ms{1}R)~~.$
\end{mpdisplay}
Applying  (\ref{Q.newdef}),  the definition of $Q$    (and using the symmetry of set union),  this proves (\ref{Pcupa0}).    
Turning now to (\ref{Pcupa1}),  we have:
\begin{mpdisplay}{0.15em}{6.5mm}{0mm}{2}
	$Q\ms{1}{\MPcomp}\ms{1}(P\ms{1}{\cup}\ms{1}R)$\push\-\\
	$=$	\>	\>$\{$	\>\+\+\+definition:  (\ref{Q.newdef})\-\-$~~~ \}$\pop\\
	$(P\ms{2}{\cup}\ms{2}R\ms{2}{\cup}\ms{2}P{\MPcomp}R\ms{2}{\cup}\ms{2}R{\MPcomp}P)\ms{1}{\MPcomp}\ms{1}(P\ms{1}{\cup}\ms{1}R)$\push\-\\
	$=$	\>	\>$\{$	\>\+\+\+distributivity and  $P$ and $R$ are pers  (so $P{\MPcomp}P\ms{1}{=}\ms{1}P$ and $R{\MPcomp}R\ms{1}{=}\ms{1}R$)\-\-$~~~ \}$\pop\\
	$P\ms{2}{\cup}\ms{2}R{\MPcomp}P\ms{2}{\cup}\ms{2}P{\MPcomp}R\ms{2}{\cup}\ms{2}R\ms{2}{\cup}\ms{2}P{\MPcomp}R{\MPcomp}P\ms{2}{\cup}\ms{2}R{\MPcomp}P{\MPcomp}R$\push\-\\
	$=$	\>	\>$\{$	\>\+\+\+lemmas \ref{PRP} and \ref{edRTR}\-\-$~~~ \}$\pop\\
	$P\ms{2}{\cup}\ms{2}R{\MPcomp}P\ms{2}{\cup}\ms{2}P{\MPcomp}R\ms{2}{\cup}\ms{2}R$\push\-\\
	$=$	\>	\>$\{$	\>\+\+\+definition:  (\ref{Q.newdef})\-\-$~~~ \}$\pop\\
	$Q~~.$
\end{mpdisplay}
\vspace{-7mm}
\MPendBox

\begin{Corollary}\label{Qtransitive}{\rm \ \ \ $Q$ is the least transitive relation that includes $P\ms{1}{\cup}\ms{1}R$.  That is,  $Q$ is the
transitive closure of $P\ms{1}{\cup}\ms{1}R$.  Also,  $Q$ is symmetric and, hence, a per.
}%
\end{Corollary}%
{\bf Proof}~~~First we show that $Q$ is transitive:
\begin{mpdisplay}{0.15em}{6.5mm}{0mm}{2}
	$Q{\MPcomp}Q$\push\-\\
	$=$	\>	\>$\{$	\>\+\+\+(\ref{Pcupa0})\-\-$~~~ \}$\pop\\
	$Q\ms{1}{\MPcomp}\ms{1}(P\ms{1}{\cup}\ms{1}R)\ms{1}{\MPcomp}\ms{1}(P\ms{1}{\cup}\ms{1}R)$\push\-\\
	$=$	\>	\>$\{$	\>\+\+\+(\ref{Pcupa1})  (applied twice)\-\-$~~~ \}$\pop\\
	$Q~~.$
\end{mpdisplay}
Now we show that it is least among all such relations: we have, for all $X$,
\begin{mpdisplay}{0.15em}{6.5mm}{0mm}{2}
	$P\ms{1}{\cup}\ms{1}R\ms{1}{\subseteq}\ms{1}X\ms{4}{\wedge}\ms{4}X{\MPcomp}X\ms{1}{\subseteq}\ms{1}X$\push\-\\
	$\Rightarrow$	\>	\>$\{$	\>\+\+\+monotonicity and transitivity of ${\subseteq}$\-\-$~~~ \}$\pop\\
	$(P\ms{1}{\cup}\ms{1}R){\MPcomp}(P\ms{1}{\cup}\ms{1}R)\ms{1}{\subseteq}\ms{1}X$\push\-\\
	$=$	\>	\>$\{$	\>\+\+\+lemma \ref{Pcupa}\-\-$~~~ \}$\pop\\
	$Q\ms{1}{\subseteq}\ms{1}X~~.$
\end{mpdisplay}
That is,  $Q$ is the transitive closure of $P\ms{1}{\cup}\ms{1}R$.

Symmetry of $Q$ is obvious from (\ref{Pcupa0}) (specifically,   $Q\ms{2}{=}\ms{2}(P\ms{1}{\cup}\ms{1}R)\ms{1}{\MPcomp}\ms{1}(P\ms{1}{\cup}\ms{1}R)$)  and the fact that $P$ and $R$ are both pers.
\MPendBox 

Corollary \ref{Qtransitive} is not used directly  in its entirety, only transitivity and symmetry 
being explicitly invoked. It is
included in order to provide further justification for the definition of $Q.$  Specifically, $Q$ is the least
per that includes both $P$ and $R$.  That it includes both $P$ and $R$ means that all the points in the index  $J$  
are combined with the points in   the equivalence class in $P$ defined by  the point  in  $P{\MPperdomain}$ chosen by
 $J$; that it is least means that other equivalence classes of $P$ are unaffected.

\subsubsection{Proof of Theorem \ref{not.maximal}}\label{ProofMainTheorem}

We are now in a position to prove the two properties of $Q$ required by theorem \ref{not.maximal}.  See lemmas 
\ref{Jules5} and \ref{Jules6} below.
\begin{Lemma}\label{Jules5}{\rm \ \ \ \begin{displaymath}P\ms{1}{\preceq}\ms{1}Q~~.\end{displaymath}
}%
\end{Lemma}%
{\bf Proof}~~~Recalling definition \ref{itt.def}, we have to prove two properties.  First,
\begin{mpdisplay}{0.15em}{6.5mm}{0mm}{2}
	$P{\MPperdomain}\ms{1}{\MPcomp}\ms{1}Q\ms{1}{\MPcomp}\ms{1}P{\MPperdomain}$\push\-\\
	$=$	\>	\>$\{$	\>\+\+\+(\ref{Q.newdef})\-\-$~~~ \}$\pop\\
	$P{\MPperdomain}\ms{1}{\MPcomp}\ms{1}(P\ms{2}{\cup}\ms{2}R\ms{2}{\cup}\ms{2}P{\MPcomp}R\ms{2}{\cup}\ms{2}R{\MPcomp}P)\ms{1}{\MPcomp}\ms{1}P{\MPperdomain}$\push\-\\
	$=$	\>	\>$\{$	\>\+\+\+distributivity and domains (specifically $\left[\ms{3}P{\MPperdomain}\ms{1}{\MPcomp}\ms{1}P\ms{2}{=}\ms{2}P\ms{2}{=}\ms{2}P\ms{1}{\MPcomp}\ms{1}P{\MPperdomain}\ms{3}\right]$)\-\-$~~~ \}$\pop\\
	$P\ms{3}{\cup}\ms{3}P{\MPperdomain}\ms{1}{\MPcomp}\ms{1}R\ms{2}{\MPcomp}\ms{2}P{\MPperdomain}\ms{3}{\cup}\ms{3}P\ms{1}{\MPcomp}\ms{1}R\ms{1}{\MPcomp}\ms{1}P{\MPperdomain}\ms{3}{\cup}\ms{3}P{\MPperdomain}\ms{1}{\MPcomp}\ms{1}R\ms{1}{\MPcomp}\ms{1}P$\push\-\\
	$=$	\>	\>$\{$	\>\+\+\+$P{\MPperdomain}\ms{1}{\subseteq}\ms{1}P$ and lemma \ref{PRP}\-\-$~~~ \}$\pop\\
	$P~~.$
\end{mpdisplay}
Second,
\begin{mpdisplay}{0.15em}{6.5mm}{0mm}{2}
	$Q\ms{1}{\MPcomp}\ms{1}P{\MPperdomain}\ms{1}{\MPcomp}\ms{1}Q$\push\-\\
	$=$	\>	\>$\{$	\>\+\+\+(\ref{Pcupa1}) and, hence, by symmetry,  $(P\ms{1}{\cup}\ms{1}R)\ms{1}{\MPcomp}\ms{1}Q\ms{3}{=}\ms{3}Q$\-\-$~~~ \}$\pop\\
	$Q\ms{2}{\MPcomp}\ms{2}(P\ms{1}{\cup}\ms{1}R)\ms{1}{\MPcomp}\ms{1}P{\MPperdomain}\ms{1}{\MPcomp}\ms{1}(P\ms{1}{\cup}\ms{1}R)\ms{2}{\MPcomp}\ms{2}Q$\push\-\\
	$=$	\>	\>$\{$	\>\+\+\+(\ref{Pcupa0})\-\-$~~~ \}$\pop\\
	$Q{\MPcomp}Q{\MPcomp}Q$\push\-\\
	$=$	\>	\>$\{$	\>\+\+\+ $Q$ is a per (corollary \ref{Qtransitive}), so $Q{\MPcomp}Q$\-\-$~~~ \}$\pop\\
	$Q~~.$
\end{mpdisplay}
\vspace{-7mm}
\MPendBox

\begin{Lemma}\label{Jules6}{\rm \ \ \ \begin{displaymath}\mathbb{I}\ms{2}{\cap}\ms{2}{\MPplattop}{\MPcomp}Q{\MPcomp}{\MPplattop}\ms{4}{\subseteq}\ms{4}Q~~.\end{displaymath}
}%
\end{Lemma}%
{\bf Proof}~~~
\begin{mpdisplay}{0.15em}{6.5mm}{0mm}{2}
	$\mathbb{I}\ms{2}{\cap}\ms{2}{\MPplattop}{\MPcomp}Q{\MPcomp}{\MPplattop}$\push\-\\
	$=$	\>	\>$\{$	\>\+\+\+lemma \ref{TopRTop.is.TopPTop}\-\-$~~~ \}$\pop\\
	$\mathbb{I}\ms{3}{\cap}\ms{3}{\MPplattop}{\MPcomp}P{\MPcomp}{\MPplattop}$\push\-\\
	$=$	\>	\>$\{$	\>\+\+\+definition:  (\ref{def.q})\-\-$~~~ \}$\pop\\
	$q$\push\-\\
	$\subseteq$	\>	\>$\{$	\>\+\+\+lemma \ref{Jules(8)}\-\-$~~~ \}$\pop\\
	$P{\MPperdomain}\ms{2}{\cup}\ms{2}J$\push\-\\
	$=$	\>	\>$\{$	\>\+\+\+definition of $R$:  (\ref{def.R}),   lemma \ref{not.maximal.lemma0} with $P\ms{1}{:=}\ms{1}R$\-\-$~~~ \}$\pop\\
	$P{\MPperdomain}\ms{2}{\cup}\ms{2}R{\MPperdomain}$\push\-\\
	$\subseteq$	\>	\>$\{$	\>\+\+\+$\left[\ms{2}P{\MPperdomain}\ms{1}{\subseteq}\ms{1}P\ms{2}\right]$ with $P\ms{1}{:=}\ms{1}P$ and $P\ms{1}{:=}\ms{1}R$; monotonicity\-\-$~~~ \}$\pop\\
	$P\ms{1}{\cup}\ms{1}R$\push\-\\
	$\subseteq$	\>	\>$\{$	\>\+\+\+definition:  (\ref{Q.newdef}),  and $\left[\ms{2}U\ms{1}{\subseteq}\ms{1}U{\cup}V\ms{2}\right]$\-\-$~~~ \}$\pop\\
	$Q~~.$
\end{mpdisplay}
 \vspace{-7mm}
\MPendBox

Lemmas \ref{Jules5} and \ref{Jules6} conclude the proof of theorem \ref{not.maximal}.
\begin{Theorem}\label{maximal.char}{\rm \ \ \ A   per $P$ is maximal 
with respect to the \textsf{thins} ordering  iff $\mathbb{I}\ms{1}{\cap}\ms{1}{\MPplattop}{\MPcomp}P{\MPcomp}{\MPplattop}\ms{2}{\subseteq}\ms{2}P$.
}
\end{Theorem}
{\bf Proof}~~~We have shown (lemma \ref{Jules.if}) that $P$ is maximal if  $\mathbb{I}\ms{1}{\cap}\ms{1}{\MPplattop}{\MPcomp}P{\MPcomp}{\MPplattop}\ms{2}{\subseteq}\ms{2}P$.  Only-if follows from
theorem \ref{not.maximal}.  Specifically, suppose $P$ is a per.  By theorem \ref{not.maximal}, 
there is a per  $Q$ such that $\mathbb{I}\ms{1}{\cap}\ms{1}{\MPplattop}{\MPcomp}Q{\MPcomp}{\MPplattop}\ms{2}{\subseteq}\ms{2}Q$ and  $P\ms{1}{\preceq}\ms{1}Q$.  So, by definition of  maximal,  if $P$ is maximal,  $P\ms{1}{=}\ms{1}Q$.  
That is, by Leibniz's rule, if $P$ is maximal,   $\mathbb{I}\ms{1}{\cap}\ms{1}{\MPplattop}{\MPcomp}P{\MPcomp}{\MPplattop}\ms{2}{\subseteq}\ms{2}P$.
\MPendBox

It is remarkable that the proof of theorem \ref{maximal.char} does not rely in any way on saturation
properties of the lattices of coreflexives or of relations in general.   The proof is entirely ``point-free''.
 Equally   remarkable is that nowhere do we use complements. 

It is also worth emphasising that we have avoided the use of the cone rule and, in so doing, have avoided a
case analysis in the statement of theorem \ref{maximal.char}.   This means that the theorem is also
applicable for non-unary relation algebras.     For concrete  relations (where the cone rule does apply), the interpretation of
theorem \ref{maximal.char} is that a per $P$ is maximal iff it is empty or is an equivalence relation.  (This is
because, by applying the cone rule, the property $\mathbb{I}\ms{1}{\cap}\ms{1}{\MPplattop}{\MPcomp}P{\MPcomp}{\MPplattop}\ms{2}{\subseteq}\ms{2}P$ simplifies to $P\ms{1}{=}\ms{1}{\MPplatbottom}\ms{2}{\vee}\ms{2}I\ms{1}{\subseteq}\ms{1}P$.)

\section{Extending \textsf{thins} to arbitrary relations}\label{General.thins.relation}

In this section, we extend the  \textsf{thins} ordering  to arbitrary relations.  The section is concluded by 
theorem  \ref{minimal.is.core} which states that the minimal elements of the extended ordering are exactly
the core relations introduced in \cite{VB2022,VB2023a}.  

Recall that $R{\MPperldom{}}$ denotes the left per-domain of $R$ and $R{\MPperrdom{}}$ denotes its right per-domain.  

\begin{Definition}[Thins]\label{itt.gen.def}{\rm \ \ \ For arbitrary relations $R$ and $S$ of the same type, the relation $R\ms{1}{\preceq}\ms{1}S$ is 
defined by\begin{displaymath}R\ms{1}{\preceq}\ms{1}S\ms{9}{\equiv}\ms{9}R{\MPperldom{}}\ms{1}{\preceq}\ms{1}S{\MPperldom{}}\ms{4}{\wedge}\ms{4}R{\MPperrdom{}}\ms{1}{\preceq}\ms{1}S{\MPperrdom{}}\ms{4}{\wedge}\ms{4}R\ms{2}{=}\ms{2}R{\MPldom{}}\ms{1}{\MPcomp}\ms{1}S\ms{1}{\MPcomp}\ms{1}R{\MPrdom{}}~~.\end{displaymath}\vspace{-9mm}
}
\MPendBox\end{Definition}

The symbol ``${\preceq}$'' is overloaded in definition \ref{itt.gen.def}.  
If $R$ and $S$ have type $A\ms{1}{\sim}\ms{1}B$, the leftmost occurrence is a 
relation on relations of type $A\ms{1}{\sim}\ms{1}B$, the middle occurrence is a relation on pers of type $A$ and the rightmost
occurrence is a relation on pers of type $B$.  
\begin{Lemma}\label{itt.gen.ord}{\rm \ \ \ The \textsf{thins} relation on arbitrary relations is an ordering relation. 
}%
\end{Lemma}%
{\bf Proof}~~~
The \textsf{thins} relation on arbitrary relations is clearly reflexive.  Transitivity is also easy to prove.  
Anti-symmetry is proven below.
\begin{mpdisplay}{0.15em}{6.5mm}{0mm}{2}
	$R\ms{1}{\preceq}\ms{1}S\ms{4}{\wedge}\ms{4}S\ms{1}{\preceq}\ms{1}R$\push\-\\
	$\Rightarrow$	\>	\>$\{$	\>\+\+\+definition \ref{itt.gen.def} with $R{,}S\ms{1}{:=}\ms{1}R{,}S$ and $R{,}S\ms{1}{:=}\ms{1}S{,}R$\-\-$~~~ \}$\pop\\
	$R\ms{2}{=}\ms{2}R{\MPldom{}}\ms{1}{\MPcomp}\ms{1}S\ms{1}{\MPcomp}\ms{1}R{\MPrdom{}}\ms{4}{\wedge}\ms{4}S\ms{2}{=}\ms{2}S{\MPldom{}}\ms{1}{\MPcomp}\ms{1}R\ms{1}{\MPcomp}\ms{1}S{\MPrdom{}}$\push\-\\
	$\Rightarrow$	\>	\>$\{$	\>\+\+\+domains\-\-$~~~ \}$\pop\\
	$(R{\MPldom{}}\ms{1}{\subseteq}\ms{1}S{\MPldom{}}\ms{4}{\wedge}\ms{4}R{\MPrdom{}}\ms{1}{\subseteq}\ms{1}S{\MPrdom{}})\ms{3}{\wedge}\ms{3}(S{\MPldom{}}\ms{1}{\subseteq}\ms{1}R{\MPldom{}}\ms{4}{\wedge}\ms{4}S{\MPrdom{}}\ms{1}{\subseteq}\ms{1}R{\MPrdom{}})$\push\-\\
	$=$	\>	\>$\{$	\>\+\+\+rearranging and anti-symmetry of ${\subseteq}$\-\-$~~~ \}$\pop\\
	$R{\MPldom{}}\ms{2}{=}\ms{2}S{\MPldom{}}\ms{4}{\wedge}\ms{4}R{\MPrdom{}}\ms{2}{=}\ms{2}S{\MPrdom{}}~~.$
\end{mpdisplay}
So
\begin{mpdisplay}{0.15em}{6.5mm}{0mm}{2}
	$R\ms{1}{\preceq}\ms{1}S\ms{4}{\wedge}\ms{4}S\ms{1}{\preceq}\ms{1}R$\push\-\\
	$\Rightarrow$	\>	\>$\{$	\>\+\+\+definition \ref{itt.gen.def}  and above\-\-$~~~ \}$\pop\\
	$R\ms{2}{=}\ms{2}R{\MPldom{}}\ms{1}{\MPcomp}\ms{1}S\ms{1}{\MPcomp}\ms{1}R{\MPrdom{}}\ms{5}{\wedge}\ms{5}R{\MPldom{}}\ms{2}{=}\ms{2}S{\MPldom{}}\ms{5}{\wedge}\ms{5}R{\MPrdom{}}\ms{2}{=}\ms{2}S{\MPrdom{}}$\push\-\\
	$\Rightarrow$	\>	\>$\{$	\>\+\+\+Leibniz\-\-$~~~ \}$\pop\\
	$R\ms{2}{=}\ms{2}S{\MPldom{}}\ms{1}{\MPcomp}\ms{1}S\ms{1}{\MPcomp}\ms{1}S{\MPrdom{}}$\push\-\\
	$=$	\>	\>$\{$	\>\+\+\+domains\-\-$~~~ \}$\pop\\
	$R\ms{2}{=}\ms{2}S~~.$
\end{mpdisplay}
\vspace{-7mm}
\MPendBox

The definition of ``minimal'' and ``maximal'' with respect to the \textsf{thins} relation on arbitrary relations is
the same as definition \ref{minimal.maximal.def} except that the dummies in the universal quantifications
range over arbitrary relations (of appropriate type).
 
We recall the definition of a core relation \cite{VB2022,VB2023a}.
\begin{Definition}[Core Relation]\label{Core.gen}{\rm \ \ \ A relation $R$ is a \emph{core relation}  iff $R{\MPldom{}}\ms{1}{=}\ms{1}R{\MPperldom{}}$ and $R{\MPrdom{}}\ms{1}{=}\ms{1}R{\MPperrdom{}}$.
\vspace{-4mm}
}
\MPendBox\end{Definition}
 \begin{Lemma}\label{core.minimal}{\rm \ \ \ A core relation is minimal with respect to the \textsf{thins} ordering on arbitrary
relations.
}%
\end{Lemma}%
{\bf Proof}~~~Suppose $S$ is a core relation.    Then, for all $R$,
\begin{mpdisplay}{0.15em}{6.5mm}{0mm}{2}
	$R\ms{1}{\preceq}\ms{1}S$\push\-\\
	$=$	\>	\>$\{$	\>\+\+\+definition \ref{itt.gen.def}\-\-$~~~ \}$\pop\\
	$R{\MPperldom{}}\ms{1}{\preceq}\ms{1}S{\MPperldom{}}\ms{4}{\wedge}\ms{4}R{\MPperrdom{}}\ms{1}{\preceq}\ms{1}S{\MPperrdom{}}\ms{4}{\wedge}\ms{4}R\ms{2}{=}\ms{2}R{\MPldom{}}\ms{1}{\MPcomp}\ms{1}S\ms{1}{\MPcomp}\ms{1}R{\MPrdom{}}$\push\-\\
	$\Rightarrow$	\>	\>$\{$	\>\+\+\+assumption: $S$ is a core relation i.e.\  $S{\MPldom{}}\ms{1}{=}\ms{1}S{\MPperldom{}}$ and $S{\MPrdom{}}\ms{1}{=}\ms{1}S{\MPperrdom{}}$,  \\
	so, by lemma \ref{corefl.is.minimal}, $S{\MPperldom{}}$ and $S{\MPperrdom{}}$ are minimal; definition \ref{minimal.maximal.def}\-\-$~~~ \}$\pop\\
	$R{\MPperldom{}}\ms{1}{=}\ms{1}S{\MPperldom{}}\ms{4}{\wedge}\ms{4}R{\MPperrdom{}}\ms{1}{=}\ms{1}S{\MPperrdom{}}\ms{4}{\wedge}\ms{4}R\ms{2}{=}\ms{2}R{\MPldom{}}\ms{1}{\MPcomp}\ms{1}S\ms{1}{\MPcomp}\ms{1}R{\MPrdom{}}$\push\-\\
	$=$	\>	\>$\{$	\>\+\+\+$S{\MPldom{}}\ms{1}{=}\ms{1}S{\MPperldom{}}$ and $S{\MPrdom{}}\ms{1}{=}\ms{1}S{\MPperrdom{}}$, Leibniz\-\-$~~~ \}$\pop\\
	$R{\MPperldom{}}\ms{1}{=}\ms{1}S{\MPldom{}}\ms{5}{\wedge}\ms{5}R{\MPperrdom{}}\ms{1}{=}\ms{1}S{\MPrdom{}}\ms{5}{\wedge}\ms{5}R\ms{2}{=}\ms{2}R{\MPldom{}}\ms{1}{\MPcomp}\ms{1}S\ms{1}{\MPcomp}\ms{1}R{\MPrdom{}}$\push\-\\
	$\Rightarrow$	\>	\>$\{$	\>\+\+\+Leibniz\-\-$~~~ \}$\pop\\
	$(R{\MPperldom{}}){\MPldom{}}\ms{2}{=}\ms{2}(S{\MPldom{}}){\MPldom{}}\ms{6}{\wedge}\ms{6}(R{\MPperrdom{}}){\MPrdom{}}\ms{2}{=}\ms{2}(S{\MPrdom{}}){\MPrdom{}}\ms{6}{\wedge}\ms{6}R\ms{2}{=}\ms{2}R{\MPldom{}}\ms{1}{\MPcomp}\ms{1}S\ms{1}{\MPcomp}\ms{1}R{\MPrdom{}}$\push\-\\
	$=$	\>	\>$\{$	\>\+\+\+per domains and domains \\
	(specifically, for left domains:  $\left[\ms{2}(R{\MPperldom{}}){\MPldom{}}\ms{1}{=}\ms{1}R{\MPldom{}}\ms{2}\right]$,   $\left[\ms{2}(R{\MPldom{}}){\MPldom{}}\ms{1}{=}\ms{1}R{\MPldom{}}\ms{2}\right]$,  \\
	similarly for right domains)\-\-$~~~ \}$\pop\\
	$R{\MPldom{}}\ms{1}{=}\ms{1}S{\MPldom{}}\ms{4}{\wedge}\ms{4}R{\MPrdom{}}\ms{1}{=}\ms{1}S{\MPrdom{}}\ms{4}{\wedge}\ms{4}R\ms{2}{=}\ms{2}R{\MPldom{}}\ms{1}{\MPcomp}\ms{1}S\ms{1}{\MPcomp}\ms{1}R{\MPrdom{}}$\push\-\\
	$\Rightarrow$	\>	\>$\{$	\>\+\+\+Leibiz and domains\-\-$~~~ \}$\pop\\
	$R\ms{1}{=}\ms{1}S~~.$
\end{mpdisplay}
Thus, by definition,  $S$ is minimal with respect to the \textsf{thins} ordering on arbitrary
relations.
\MPendBox

For reference, we include the definition of an index of an arbitrary relation and several of its properties.  Proofs are
given in \cite{VB2022,VB2023a}.
\begin{Definition}[Index]\label{gen.index}{\rm \ \ \ An \emph{index} of a relation $R$ is a relation $J$ that  has the following properties:
\begin{description}
\item[(a)]$J\ms{1}{\subseteq}\ms{1}R~~,$ 
\item[(b)]$R{\MPperldom{}}\ms{1}{\MPcomp}\ms{1}J\ms{1}{\MPcomp}\ms{1}R{\MPperrdom{}}\ms{3}{=}\ms{3}R~~,$ 
\item[(c)]$J{\MPldom{}}\ms{1}{\MPcomp}\ms{1}R{\MPperldom{}}\ms{1}{\MPcomp}\ms{1}J{\MPldom{}}\ms{3}{=}\ms{3}J{\MPldom{}}~~,$ 
\item[(d)]$J{\MPrdom{}}\ms{1}{\MPcomp}\ms{1}R{\MPperrdom{}}\ms{1}{\MPcomp}\ms{1}J{\MPrdom{}}\ms{3}{=}\ms{3}J{\MPrdom{}}~~.$ 
 
\end{description}
\vspace{-7mm}
}
\MPendBox\end{Definition}

\begin{Lemma}\label{index.per.is.pid}{\rm \ \ \ If $J$ is an index of the  relation $R$ then  \begin{displaymath}J{\MPperldom{}}\ms{2}{\subseteq}\ms{2}R{\MPperldom{}}\ms{6}{\wedge}\ms{6}J{\MPperrdom{}}\ms{2}{\subseteq}\ms{2}R{\MPperrdom{}}~~.\end{displaymath}It follows that \begin{displaymath}J{\MPldom{}}\ms{2}{=}\ms{2}J{\MPperldom{}}\ms{6}{\wedge}\ms{6}J{\MPrdom{}}\ms{2}{=}\ms{2}J{\MPperrdom{}}~~.\end{displaymath}That is,  an index is a core relation.
\vspace{-9mm}
}%
\end{Lemma}%
\MPendBox

\begin{Lemma}\label{R-perleft}{\rm \ \ \ Suppose  $J$ is an index of $R$.  Then
\begin{description}
\item[(a)]$R{\MPperldom{}}\ms{1}{\MPcomp}\ms{1}J{\MPldom{}}\ms{1}{\MPcomp}\ms{1}R{\MPperldom{}}\ms{3}{=}\ms{3}R{\MPperldom{}}~~,$ 
\item[(b)]$R{\MPperrdom{}}\ms{1}{\MPcomp}\ms{1}J{\MPrdom{}}\ms{1}{\MPcomp}\ms{1}R{\MPperrdom{}}\ms{3}{=}\ms{3}R{\MPperrdom{}}~~.$ 
 
\end{description}
\vspace{-9mm}
}%
\end{Lemma}%
\MPendBox

\begin{Theorem}\label{index-perdoms}{\rm \ \ \ Suppose  $J$ is an index of $R$.  Then $J{\MPldom{}}$ is an index of $R{\MPperldom{}}$ and $J{\MPrdom{}}$ is an index of $R{\MPperrdom{}}$.
}
\MPendBox\end{Theorem}

We now resume the study of the extended \textsf{thins} ordering. 
\begin{Lemma}\label{index.thins.gen}{\rm \ \ \ If $J$ is an index of $R$ then $J\ms{1}{\preceq}\ms{1}R$.  
}%
\end{Lemma}%
{\bf Proof}~~~Suppose that $J$ is an index of $R$.   By definition \ref{itt.gen.def}, 
we have to prove that $J{\MPperldom{}}\ms{1}{\preceq}\ms{1}R{\MPperldom{}}$,  $J{\MPperrdom{}}\ms{1}{\preceq}\ms{1}R{\MPperrdom{}}$ and $J\ms{2}{=}\ms{2}J{\MPldom{}}\ms{1}{\MPcomp}\ms{1}R\ms{1}{\MPcomp}\ms{1}J{\MPrdom{}}$.  For the first property, we have:
\begin{mpdisplay}{0.15em}{6.5mm}{0mm}{2}
	$J{\MPperldom{}}\ms{2}{\preceq}\ms{2}R{\MPperldom{}}$\push\-\\
	$=$	\>	\>$\{$	\>\+\+\+definition \ref{itt.def}\-\-$~~~ \}$\pop\\
	$J{\MPperldom{}}\ms{3}{=}\ms{3}(J{\MPperldom{}}){\MPperdomain}\ms{1}{\MPcomp}\ms{1}R{\MPperldom{}}\ms{1}{\MPcomp}\ms{1}(J{\MPperldom{}}){\MPperdomain}\ms{7}{\wedge}\ms{7}R{\MPperldom{}}\ms{3}{=}\ms{3}R{\MPperldom{}}\ms{1}{\MPcomp}\ms{1}(J{\MPperldom{}}){\MPperdomain}\ms{1}{\MPcomp}\ms{1}R{\MPperldom{}}$\push\-\\
	$=$	\>	\>$\{$	\>\+\+\+domains (specifically $\left[\ms{2}(R{\MPperldom{}}){\MPperdomain}\ms{1}{=}\ms{1}R{\MPldom{}}\ms{2}\right]$ with $R\ms{1}{:=}\ms{1}J$)\-\-$~~~ \}$\pop\\
	$J{\MPperldom{}}\ms{3}{=}\ms{3}J{\MPldom{}}\ms{1}{\MPcomp}\ms{1}R{\MPperldom{}}\ms{1}{\MPcomp}\ms{1}J{\MPldom{}}\ms{7}{\wedge}\ms{7}R{\MPperldom{}}\ms{3}{=}\ms{3}R{\MPperldom{}}\ms{1}{\MPcomp}\ms{1}J{\MPldom{}}\ms{1}{\MPcomp}\ms{1}R{\MPperldom{}}$\push\-\\
	$=$	\>	\>$\{$	\>\+\+\+$J$ is an index of $R$:  theorem \ref{index-perdoms}  and definition \ref{gen.index}(c) with $J{,}R\ms{2}{:=}\ms{2}J{\MPldom{}}\ms{1}{,}\ms{1}R{\MPperldom{}}$;\\
	 $J$ is an index of $R$:  lemma \ref{R-perleft}(a)\-\-$~~~ \}$\pop\\
	$J{\MPperldom{}}\ms{2}{=}\ms{2}J{\MPldom{}}\ms{7}{\wedge}\ms{7}\mathsf{true}$\push\-\\
	$=$	\>	\>$\{$	\>\+\+\+lemma \ref{index.per.is.pid}\-\-$~~~ \}$\pop\\
	$\mathsf{true}~~.$
\end{mpdisplay}
By symmetry,  $J{\MPperrdom{}}\ms{1}{\preceq}\ms{1}R{\MPperrdom{}}$.    The third property is straightforward:
\begin{mpdisplay}{0.15em}{6.5mm}{0mm}{2}
	$J{\MPldom{}}\ms{1}{\MPcomp}\ms{1}R\ms{1}{\MPcomp}\ms{1}J{\MPrdom{}}$\push\-\\
	$=$	\>	\>$\{$	\>\+\+\+assumption: $J$ is an index of $R$,  definition \ref{gen.index}(b)\-\-$~~~ \}$\pop\\
	$J{\MPldom{}}\ms{1}{\MPcomp}\ms{1}R{\MPperldom{}}\ms{1}{\MPcomp}\ms{1}J\ms{1}{\MPcomp}\ms{1}R{\MPperrdom{}}\ms{1}{\MPcomp}\ms{1}J{\MPrdom{}}$\push\-\\
	$=$	\>	\>$\{$	\>\+\+\+domains\-\-$~~~ \}$\pop\\
	$J{\MPldom{}}\ms{1}{\MPcomp}\ms{1}R{\MPperldom{}}\ms{1}{\MPcomp}\ms{1}J{\MPldom{}}\ms{1}{\MPcomp}\ms{1}J\ms{1}{\MPcomp}\ms{1}J{\MPrdom{}}\ms{1}{\MPcomp}\ms{1}R{\MPperrdom{}}\ms{1}{\MPcomp}\ms{1}J{\MPrdom{}}$\push\-\\
	$=$	\>	\>$\{$	\>\+\+\+assumption: $J$ is an index of $R$,  definition \ref{gen.index}(c) and \ref{gen.index}(d)\-\-$~~~ \}$\pop\\
	$J{\MPldom{}}\ms{1}{\MPcomp}\ms{1}J\ms{1}{\MPcomp}\ms{1}J{\MPrdom{}}$\push\-\\
	$=$	\>	\>$\{$	\>\+\+\+domains\-\-$~~~ \}$\pop\\
	$J~~.$
\end{mpdisplay}
\vspace{-7mm}
\MPendBox

\begin{Lemma}\label{thins.min.index}{\rm \ \ \ If  relation $S$ is minimal with respect to the \textsf{thins} ordering on arbitrary
relations and $J$ is an index of $S$ then $J\ms{1}{=}\ms{1}S$. 
}%
\end{Lemma}%
{\bf Proof}~~~Immediate from lemma \ref{index.thins.gen} and the definition of minimal.
\MPendBox

\begin{Lemma}\label{thins.minmax.gen}{\rm \ \ \ Assuming axiom \ref{Axiom of Choice} (our axiom of choice),  if  relation $S$ is minimal with respect to 
the \textsf{thins} ordering on arbitrary relations then $S{\MPperldom{}}$ and  $S{\MPperrdom{}}$ are  mimimal with respect to the
 \textsf{thins} ordering on pers.
}%
\end{Lemma}%
{\bf Proof}~~~ 
\begin{mpdisplay}{0.15em}{6.5mm}{0mm}{2}
	$S$ is minimal\push\-\\
	$\Rightarrow$	\>	\>$\{$	\>\+\+\+lemma \ref{thins.min.index} (assuming axiom \ref{Axiom of Choice})\-\-$~~~ \}$\pop\\
	$S$ is an index of $S$\push\-\\
	$\Rightarrow$	\>	\>$\{$	\>\+\+\+lemma \ref{index.per.is.pid}  and theorem \ref{index-perdoms}\-\-$~~~ \}$\pop\\
	$S{\MPldom{}}\ms{2}{=}\ms{2}S{\MPperldom{}}\ms{6}{\wedge}\ms{6}S{\MPldom{}}$ is an index of $S{\MPperldom{}}$\push\-\\
	$\Rightarrow$	\>	\>$\{$	\>\+\+\+lemma \ref{corefl.is.minimal}\-\-$~~~ \}$\pop\\
	$S{\MPperldom{}}$ is minimal~~.
\end{mpdisplay}
Symmetrically,  if $S$ is minimal then $S{\MPperrdom{}}$ is minimal.
\MPendBox

Note that the axiom of choice is invoked  in the proof of lemma \ref{thins.minmax.gen}: the
application of lemma \ref{thins.min.index} in the first step assumes that $S$ has an index and that this is so is a
consequence of the axiom of choice \cite{VB2022,VB2023a}.   Consequently,  the axiom of choice is also
required  in the statement and proof of the main theorem of this section:
\begin{Theorem}\label{minimal.is.core}{\rm \ \ \ Assuming axiom \ref{Axiom of Choice} (our axiom of choice), 
a relation $S$ is minimal with respect to the  \textsf{thins} ordering  on arbitrary  relations iff $S$ is a core relation.
}
\end{Theorem}
{\bf Proof}~~~``If'' is lemma \ref{core.minimal}.  ``Only if'' is a combination of lemma \ref{thins.minmax.gen} and
theorem \ref{choice.defs.minimal}:
\begin{mpdisplay}{0.15em}{6.5mm}{0mm}{2}
	$S$ is minimal\push\-\\
	$\Rightarrow$	\>	\>$\{$	\>\+\+\+lemma \ref{thins.minmax.gen} (assuming axiom \ref{Axiom of Choice})\-\-$~~~ \}$\pop\\
	$S{\MPperldom{}}$ is minimal\push\-\\
	$\Rightarrow$	\>	\>$\{$	\>\+\+\+theorem  \ref{choice.defs.minimal}\-\-$~~~ \}$\pop\\
	$S{\MPperldom{}}\ms{1}{\subseteq}\ms{1}\mathbb{I}$\push\-\\
	$=$	\>	\>$\{$	\>\+\+\+domains\-\-$~~~ \}$\pop\\
	$(S{\MPperldom{}}){\MPldom{}}\ms{2}{=}\ms{2}S{\MPperldom{}}$\push\-\\
	$=$	\>	\>$\{$	\>\+\+\+per domains\-\-$~~~ \}$\pop\\
	$S{\MPldom{}}\ms{2}{=}\ms{2}S{\MPperldom{}}~~.$
\end{mpdisplay}
That is,\begin{displaymath}S\mbox{ is minimal\ms{6}}{\Rightarrow}\ms{6}S{\MPldom{}}\ms{2}{=}\ms{2}S{\MPperldom{}}~~.\end{displaymath}Symmetrically,  \begin{displaymath}S\mbox{ is minimal\ms{6}}{\Rightarrow}\ms{6}S{\MPrdom{}}\ms{2}{=}\ms{2}S{\MPperrdom{}}~~.\end{displaymath}Thus, by definition \ref{Core.gen},  \begin{displaymath}S\mbox{ is minimal\ms{4}}{\Rightarrow}\ms{4}S\mbox{ is a core relation}~~.\end{displaymath}\vspace{-7mm}
\MPendBox

\section{Conclusion}\label{itt:Conclusion}

For us, the primary purpose of point-free relation algebra is to enable precise and concise reasoning
about  binary relations.  Therefore,  the axiomatisation does not capture all the properties of concrete
relations and sometimes it is necessary to add axioms in order to facilitate such reasoning.    Earlier work
\cite{VB2022,VB2023a} focused on facilitating pointwise reasoning ---whilst not compromising the
concision and precision of point-free reasoning---  by adding axioms expressing the powerset properties
of relations of a given type.  To this end, \cite{VB2022,VB2023a} proposed an  axiom of choice
(axiom \ref{Axiom of Choice}) together with a saturation axiom (the axiom that the lattice of coreflexives of
a given type is saturated by points).  In this paper, the focus is  on just the axiom of choice.  

A central contribution is  to provide further insight into the notion of a core relation introduced in 
\cite{VB2022,VB2023a}.  Theorem \ref{minimal.is.core} shows that the relations that are minimal with respect to
the \textsf{thins} relation are precisely the core relations.  

The most challenging aspect of the paper has been the proof of 
theorem \ref{maximal.char}, which characterises pers that are maximal with respect to the \textsf{thins} ordering.
In meeting this challenge, a major contribution is  the introduction of  a new idiom  to  point-free
relation algebra that avoids the case analysis on whether or not a relation is empty.  In formal terms,  we
avoid appeals to the cone rule.    As a consequence, 
we extend the validity of the characterisation to models quite different from concrete relations. 
It remains to be seen whether or not this will be beneficial in practical applications.

The  challenge we imposed on ourselves  has undoubtedly increased the length of the proof  of theorem 
\ref{maximal.char}  considerably:  several  of the lemmas  (for
example, lemma  \ref{TopRTop.is.TopPTop}) are trivial if the cone rule is assumed.  The fact that we have
overcome the challenge attests to the strength of our axiom of choice.

\paragraph{Acknowledgement}\label{Acknowledgement.thins.relation}

Many thanks to Michael Winter for valuable suggestions in the early stages of this work.

\bibliographystyle{alpha}
\bibliography{bibliogr}

\end{document}